\documentclass[sn-nature]{sn-jnl}% Standard Nature Portfolio Reference Style

\usepackage{times}

\usepackage[utf8]{inputenc}
\usepackage{tikz}
\usetikzlibrary{positioning,shapes.geometric, arrows,calc}

\tikzstyle{process} = [rectangle, draw = black, text width= 5cm, minimum height=1cm, align = left]
\tikzstyle{decision} = [rectangle, text width=5cm, minimum height=1cm, align = left, draw=black]
\tikzstyle{arrow} = [thick,->,>=stealth]
\tikzstyle{line} = [thick,-]

\usepackage{graphicx}
\usepackage{caption}
\usepackage{floatpag}
\usepackage{subfloat}
\usepackage{subcaption}

\usepackage[dvipsnames]{xcolor}

\usepackage{amsmath}
\usepackage{xfrac}

\usepackage{longtable}
\usepackage{booktabs,caption,fixltx2e}
\usepackage{threeparttablex,threeparttable}
\usepackage{array}
\usepackage{pdflscape}
\usepackage{enumitem}
\usepackage{afterpage}
\usepackage{siunitx}

\newcommand{\circled}[1]{\tikz[baseline=(char.base)]{\node[shape=circle,draw,inner sep=2pt] (char) {#1};}}

\newcolumntype{L}[1]{>{\raggedright\let\newline\\\arraybackslash\hspace{0pt}}p{#1}}
\newcolumntype{C}[1]{>{\centering\let\newline\\\arraybackslash\hspace{0pt}}p{#1}}
\newcolumntype{R}[1]{>{\raggedleft\let\newline\\\arraybackslash\hspace{0pt}}p{#1}}

\begin{document}

\title[A robot-assisted scanning pipeline]{A robot-assisted pipeline to rapidly scan 1.7 million historical aerial photographs}

\author[1]{\fnm{Sheila} \sur{Masson}}
\author[1]{\fnm{Alan} \sur{Potts}}
\author[1,2]{\fnm{Allan} \sur{Williams}}
\author[3]{\fnm{Steve} \sur{Berggreen}}
\author[1]{\fnm{Kevin} \sur{McLaren}}
\author[1]{\fnm{Sam} \sur{Martin}}
\author[3]{\fnm{Eugenio} \sur{Noda}}
\author[4]{\fnm{Nicklas} \sur{Nordfors}}
\author[1]{\fnm{Nic} \sur{Ruecroft}}
\author[5,6]{\fnm{Hannah} \sur{Druckenmiller}}
\author[3,6]{\fnm{Solomon} \sur{Hsiang}}
\author[7,8]{\fnm{Andreas} \sur{Madestam}}
\author*[9,10]{\fnm{Anna} \sur{Tompsett}\email{anna.tompsett@iies.su.se}}

\affil[1]{\orgname{National Collection of Aerial Photography, Historic Environment Scotland}, \orgaddress{\city{Edinburgh}, \country{UK}}}

\affil[2]{\orgname{Center for Geospatial Intelligence, George Mason University}, \orgaddress{\city{Fairfax}, \country{USA}}}

\affil[3]{\orgname{Global Policy Laboratory, Stanford Doerr School of Sustainability, Stanford University}, \orgaddress{\city{Palo Alto}, \country{USA}}}

\affil[4]{\orgdiv{Department of Economics}, \orgname{University of Gothenburg}, \orgaddress{\city{Gothenburg}, \country{Sweden}}}

\affil[5]{\orgdiv{Division of Humanities and Social Sciences}, \orgname{California Institute of Technology}, \orgaddress{\city{Pasadena}, \country{USA}}}

\affil[6]{\orgname{National Bureau of Economic Research}, \orgaddress{\city{Cambridge}, \country{USA}}}

\affil[7]{\orgdiv{Department of Economics}, \orgname{Stockholm University}, \orgaddress{\city{Stockholm}, \country{Sweden}}}

\affil[8]{\orgname{Centre for Economic Policy Research}, \country{UK}}

\affil[9]{\orgname{Beijer Institute of Ecological Economics, Royal Swedish Academy of Sciences}, \orgaddress{\city{Stockholm}, \country{Sweden}}}

\affil[10]{\orgname{Institute for International Economic Studies, Stockholm University}, \orgaddress{\city{Stockholm}, \country{Sweden}}}

\date{}

\abstract{During the 20th Century, aerial surveys captured hundreds of millions of high-resolution photographs of the earth's surface. These images, the precursors to modern satellite imagery, represent an extraordinary visual record of the environmental and social upheavals of the 20th Century. However, most of these images currently languish in physical archives where retrieval is difficult and costly. Digitization could revolutionize access, but manual scanning is slow and expensive. Automated scanning could make at-scale digitization feasible, unlocking this visual record of the 20th Century for the digital era. Here, we describe and validate a novel robot-assisted pipeline that increases worker productivity in scanning 30-fold, applied at scale to digitize an archive of 1.7 million historical aerial photographs from 65 countries.  }

\maketitle

%\clearpage

\section*{Introduction}

Aerial photography revolutionized the speed and cost of cartography in the 20th Century. A single flight and a week of laboratory processing could map the same area as a team of ground surveyors in a whole year \cite{kinney1919broad}. Around the world, pilots logged millions of flight hours to capture hundreds of millions of aerial photographs of the earth's surface. While some of these photographs were lost or destroyed, a survey reveals that more than 140 million have been carefully preserved in dozens of archives worldwide (Tab \ref{tab:archives}). This extraordinary visual record holds the potential to revolutionize our understanding of the 20th Century, just as modern satellite imagery has transformed our understanding of the world in the 21st Century \cite{burke2021using}. With only relatively small historical aerial photography datasets, scientists have tracked glacier change \cite{domgaard2024early}, shed new light on the impacts of sea level rise \cite{webb2010dynamic}, rediscovered indigenous land management techniques \cite{blackwood2022pirra}, and learned about how to prevent deforestation \cite{andam2008measuring}. 

Access to physical archives remains extremely limited, however. Storing a million photographic prints requires approximately 1.4km of shelf space. Most of the larger aerial photography archives are thus kept in off-site storage, making it costly to search for, extract, and view any images. The negatives and prints in these archives are also, inevitably, ageing and fading \cite{lavedrine2009photographs}. Storage in conditions that are too warm or too humid hastens the rate of decay \cite{lavedrine2009photographs}. To unlock the data stored within these archives for widespread use, and to ensure the preservation of these archives for future generations, many archives are seeking to create digital copies of the physical images. Available in digital form at scale, historical aerial photographs could open the door to an entirely new quantitative understanding of the extraordinary environmental and social changes that characterized the 20th Century. 

%Iron Mountain boxes: D400mm x W350mm x H265mm 
%2 boxes require 1 m of shelf space. 4,679 cartons/1.7 million * 0.5m = 1.4 km of shelf space.  

The manual approach to scanning is, however, prohibitively costly and time-consuming. As a result, although most archives have begun to digitize their physical archives, few approach completion \cite{giordano2019archiving}. Here, we describe a robot-assisted pipeline that increases scanning rates per full-time worker more than thirty-fold without compromising the physical integrity of the historical photographs, applied to an archive of 1.7 million historical aerial photographs covering 65 countries worldwide.  

%65 countries is based on the original NCAP catalogue with placenames from boxes, matching to modern country boundaries. 

\section*{Materials and Methods}

\subsection*{Images of a third of the world's countries throughout the 20th Century} 

We developed the pipeline to digitize an archive of 1.7 million images, held and curated since 2012 by the National Collection of Aerial Photography (NCAP) in Edinburgh, UK (part of Historic Environment Scotland). The archive was created by a government unit, originally part of the Colonial Office, established in 1946 as the Directorate of Colonial Surveys. One of its central objectives was to produce accurate topographical maps of the ``Colonial Empire'' \cite{macdonald1996mapping}. The unit outlived the decline of the British Empire, being renamed the Directorate of Overseas Surveys (DOS) and widening its remit to developing countries more broadly. The DOS survived in some form---eventually incorporated into the the Ordnance Survey, the British national mapping agency---until 2001 \cite{macdonald1996mapping}, well into the era of modern satellite imagery.

Over its history, the DOS mapped more than 2 million square miles (more than 5 million km$^2$) of the earth's land surface \cite{macdonald1996mapping}. To produce these maps, the organization commissioned aerial surveys, building on advances in aerial photogrammetry during the First and Second World Wars \cite{macdonald1996mapping}. The first million square miles of coverage were flown by the British Royal Air Force (RAF) \cite{macdonald1996mapping}. The earliest RAF-flown missions used the unforgiving K17 Fairchild camera, which required a vacuum to flatten film before exposure and produced inadequate quality images when handled with insufficient care, later superseded by the more reliable Williamson F49 \cite{macdonald1996mapping,monmonier2019history}. Later surveys were procured commercially from independent firms, who used a wide variety of aircraft and cameras. 

Each survey or contract set out to cover an entire country or region in a particular window of time. To complete a survey, reconnaissance aircraft flew sorties or missions, in this context to photograph a specific area for mapping, recording images on film. In the earliest surveys, in the absence of any geographical information with which to navigate, pilots flew in concentric circles around radar beacons \cite{macdonald1996mapping}. Later, as navigation instrumentation improved, pilots flew in long parallel swathes, known anecdotally as ``mowing the lawn''. 

The exact geographical coverage of images in the DOS archive is difficult to confirm. Matching placenames on box labels to modern day countries yields the estimate of 65 countries. For an exact coverage map, one would need to georeference each image in the archive. To do so manually for the entire DOS archive would require approximately 60 person-years. Automated georeferencing processes are still under development \cite{noda2024machine}, but we obtain a preliminary coverage estimate by digitizing maps of photographic cover from DOS annual reports (Fig. \ref{fig:map}). To the extent that not all photographed areas were mapped in annual reports, the map understates the true extent covered by the archive.  

In the process, the organization accumulated an extensive and meticulously curated archive, including two libraries of photographic prints: a master copy and a second, working copy. As the unit wound down and finally closed, the archive was dispersed. While film negatives were transferred to the relevant national governments or mapping agencies of the covered countries, the master print library was entrusted to the British Empire and Commonwealth Museum in Bristol, England \cite{hadley2006changes}. Unfortunately, however, storage conditions at the museum proved damp and inadequate, and when the museum later closed---due to financial difficulties \cite{pucknell2020british} and amid scandal over unauthorized sales of objects \cite{harris2011rise}---the archive was on the verge of being destroyed. NCAP secured permission to take custody of the archive and transferred it to controlled and consistent storage conditions.

\afterpage{%
\begin{landscape}
\begin{figure}[p]
    \thisfloatpagestyle{empty}
    \centerline{
    \includegraphics[width=1.1\linewidth]{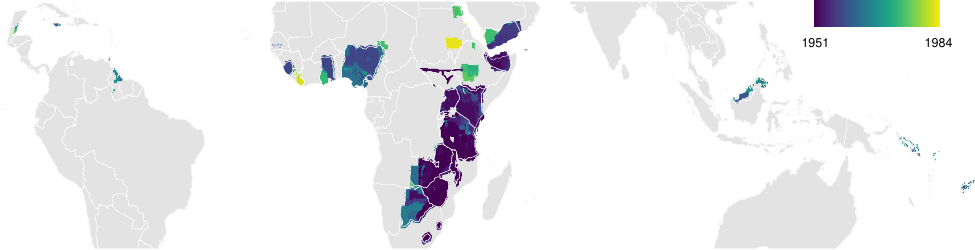}

    }
     \captionsetup{margin={0cm,-1cm}}

     \caption{\raggedright \textbf{Approximate coverage of the photographs in the DOS archive} Estimated based on digitization of coverage maps that were published in DOS Annual Reports between 1951 and 1984. Shading indicates the approximate year of earliest coverage in the archive. Coverage from between 1946 and 1951 is shown as from 1951, based on the earliest available coverage map, published in 1951 and showing coverage to that date. Between 1951 and 1966, maps of coverage were cumulative, i.e., displaying all coverage up to and including the report year. We differenced the coverage areas between years to obtain newly covered areas. From 1967, maps showed recent coverage in a different colour, allowing us to observe recent coverage directly. Land in grayscale, modern country boundaries displayed in white. Mollweide projection. \label{fig:map}}

\end{figure}

\end{landscape}
}

All prints in the archive are contact prints, produced by laying film negatives flat on photographic paper and exposing to light. The resulting printed image has exactly the same size as the original negative. DOS surveys typically used 9x9 inch negatives, although the archive also contains prints from other negative sizes (7x9, 5x5, and 22x11.5). Prints display both the photographic image and any marginalia, including fiduciary marks and metadata recorded contemporaneously, such as the date captured, the general geographic location, the sortie, the frame number, the image type (typically vertical, when captured for cartographic purposes), and the scale (see, e.g., Fig 
\ref{fig:scan}). The prints themselves vary in shape and margin width, many having been hand cut. Most are black and white prints from panchromatic negatives; some are black and white prints from colour or colour infra-red negatives; and a few are colour prints. Prints are organized by survey or contract, and then sortie or mission. Sorties are labeled following an established standard (Tab. \ref{tab:identifiers}). 

Prints were originally stored in plain brown cardboard boxes, or, occasionally, commercial print boxes from the photographic paper manufacturer. The original boxes have printed or hand-written notations on the outside edge, detailing the country, air force or company, year, sortie code, print numbers, and other information. All boxes were photographed on accession to the NCAP collection.  

%% Notes NN:
% we have scales for 103 of 123 images
% removed vanikoro so 123
% weighting by share of total

% scale denominator:
%    wt_mean wt_median  wt_iqr      min      max
%    36739.48  42579.23 25893.6 4964.169 82283.46

% altitude in meters:
%    wt_mean wt_median   wt_iqr min   max
%   5543.611  6262.106 2586.351 762 12192

% 152 mm is approximate

Missions were flown at a range of altitudes to create maps at different scales. We randomly sampled 124 images from the archive, stratifying the sample to sample one image at random for each country and decade of coverage, if we had collated data on the year of coverage, and one image at random for each country otherwise. For each image, we recorded the focal length ($f$) and the altitude ($H$), or the scale ($S$), as shown in the metadata (see, e.g., Fig \ref{fig:metadata}). Weighting images according to the share of the archive in each country or country-decade stratum, the mean altitude in meters was 5543 (median 6262; interquartile range 5722--8308, see also Fig \ref{fig:altitudes}). Most cameras had a focal length of about 152 mm, approximately 6$^{\prime\prime}$. Approximating scale $S$ as $\sfrac{f}{H}$ where scale was not separately specified, we estimate a range of scales from about 1:5,000 to 1:80,000 (median 42,579; interquartile range 36,510--62,403). These approximations do not account for the local difference between ground level and sea level---to account for this, we would need to georeference each image, which is difficult without contextual information---so the estimated denominators are an upper bound.

Each contact print has a resolution, typically measured in line-pairs per millimeter (mm) that can be visually resolved \cite{slama1980manual}. The resolution depends on the intrinsic properties of the photographic film and paper, but also on blur and the speed of movement of the aircraft \cite{slama1980manual,clark1944photographic}. Estimates of the resolution of aerial photographic prints in practice vary between 10-14 line pairs/mm in the 1940s \cite{clark1944photographic} to 27 line pairs/mm later in the 20th Century \cite{light1993national}.

The ground resolved resolution---the size of the smallest object on the ground that can be distinguished on the print---is given by dividing the smallest resolvable feature in the print by the scale. At resolution of 10 line pairs/mm, the smallest resolvable object in a print would be 50~$\mu$m; at 27 line pairs/mm, it would be 18~$\mu$m. At the median scale in our data, these correspond to ground resolved resolutions of 2.1~m and 0.8~m, respectively.

\subsection*{A robot-assisted preservation and scanning pipeline} 

Scanning the DOS archive manually would require approximately 10 person-years of tedious, physically demanding labour, even with fast, modern scanners. But the labour required is extremely monotonous and physically demanding, and workers bear a high risk of repetitive strain injury. To make digitization at scale feasible, we developed a robot-assisted pipeline. The pipeline is an example of a cobot \cite{colgate1996cobots} or collaborative robotic \cite{goldberg2019robots} system. The pipeline leverages the nuance and judgement of human workers during the ``preservation pipeline'', in which prints are retrieved from storage and prepared for scanning, at which stage the range of tasks required are varied and complex. The pipeline then leverages the precision and untiring capacity for repetitive tasks of the robot during the ``scanning pipeline'', in which each print is scanned to create a digital copy. The physical prints are returned to storage, while the digital copies are stored on a server after post-processing (Fig. \ref{fig:overview}). 

\afterpage{
\begin{landscape}
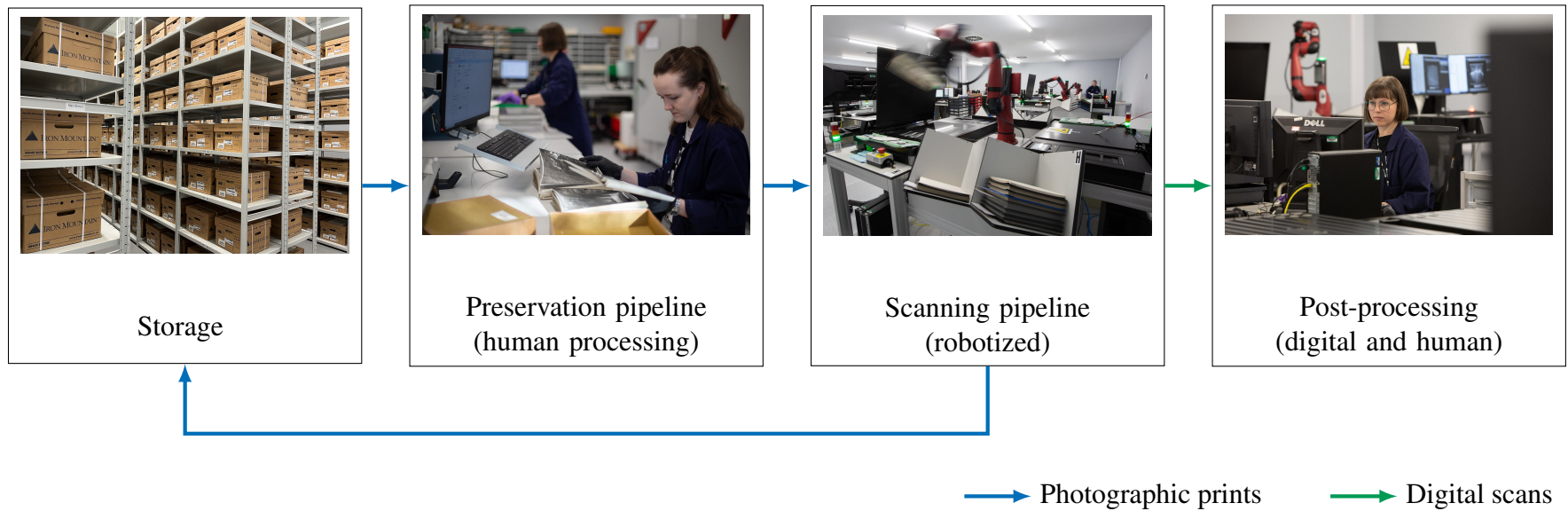
\begin{figure}[p]
\thisfloatpagestyle{empty}
\centerline{
\begin{tikzpicture}[
    node distance = 6cm, auto,
    block/.style = {rectangle, draw, text width=5cm, text centered, minimum height=4.5cm},
    line/.style = {draw, ->, >=latex, ultra thick}
]

    % Nodes
    \node [block] (storage) {
        \includegraphics[width=4.9cm]{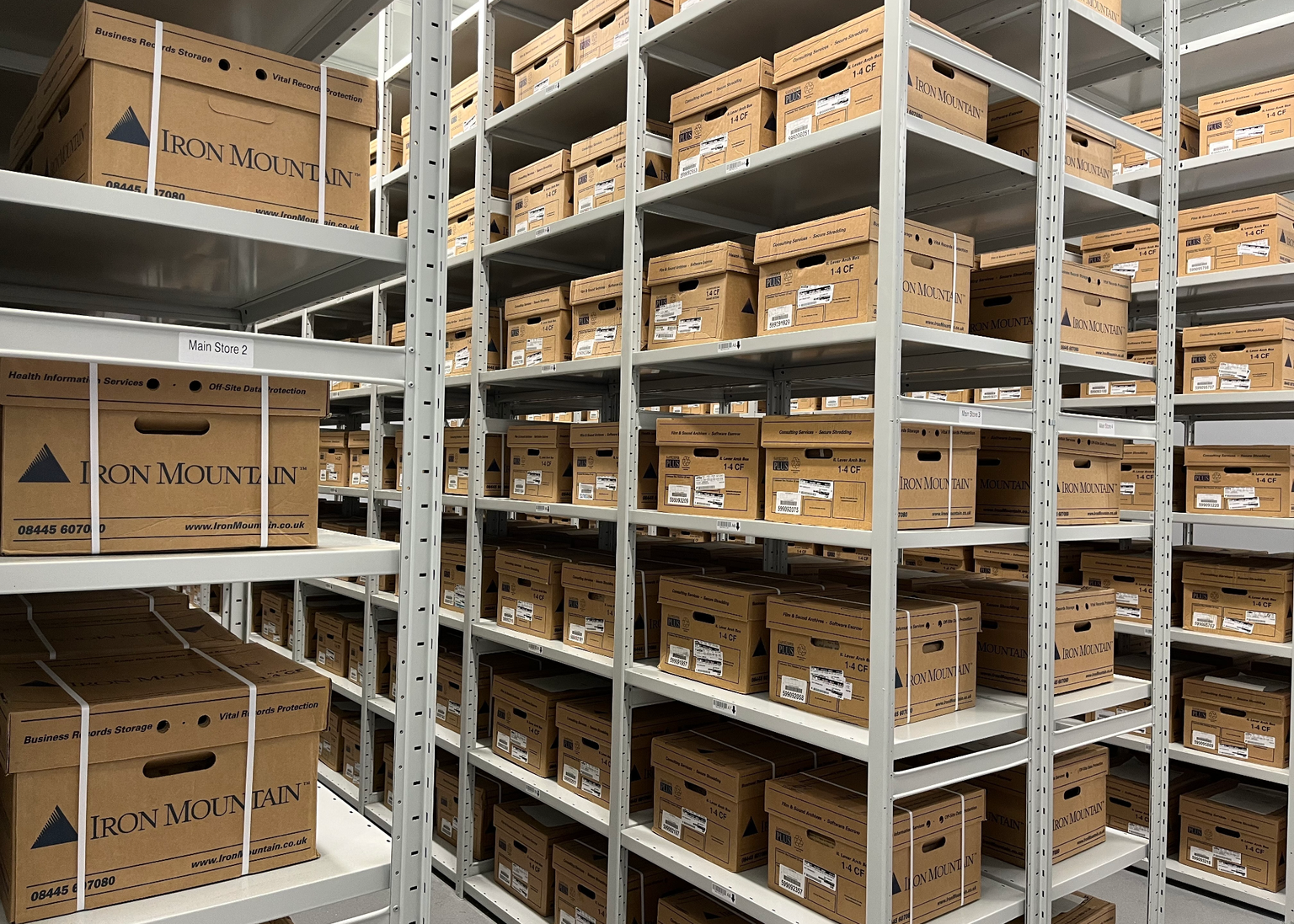}
         \vspace{1ex}\\
        Storage \vspace{1ex} \\

    };

    \node [block, right of=storage] (preservation) {
        \includegraphics[width=4.9cm]{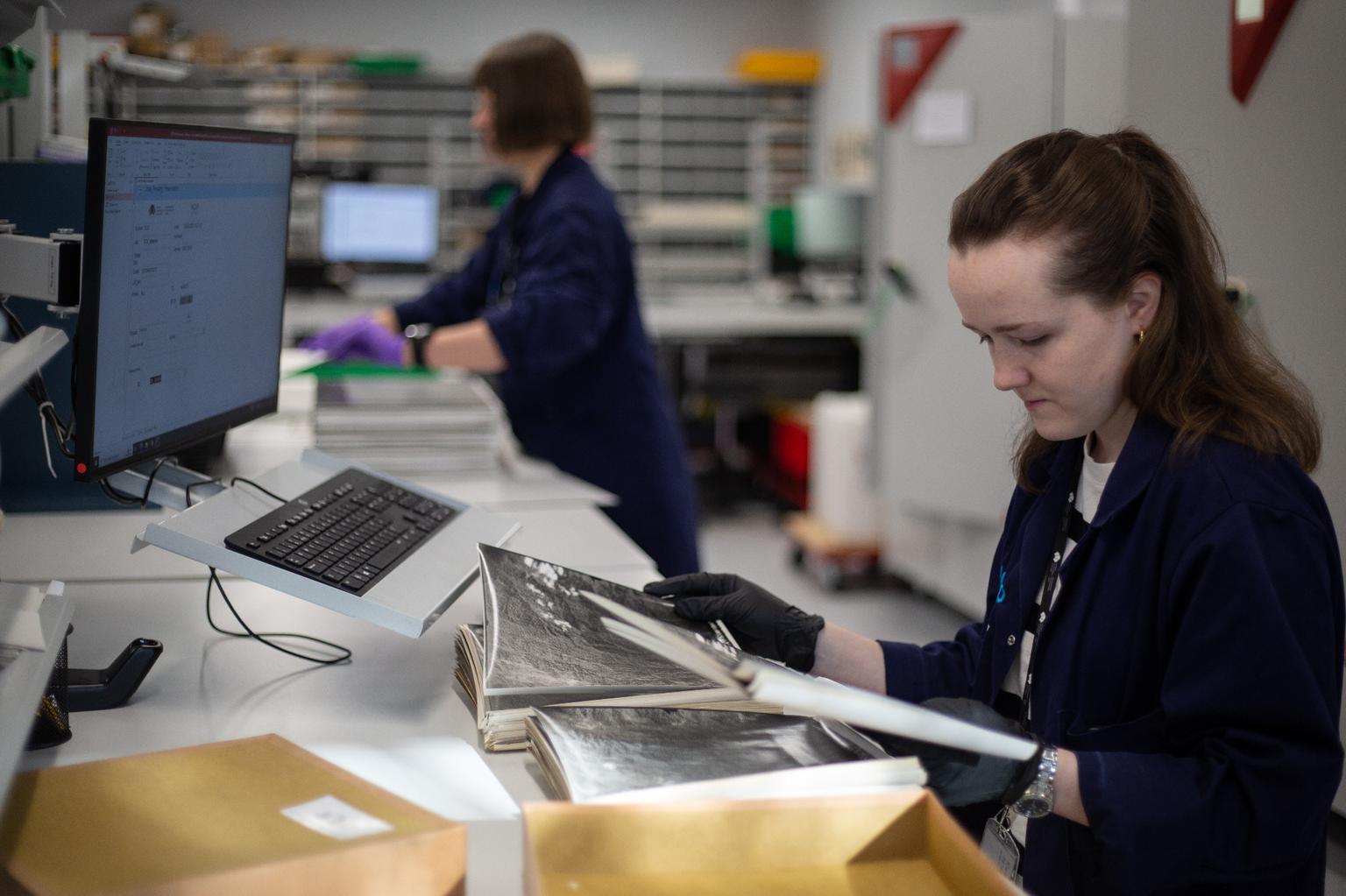}
        \vspace{1ex}\\
        Preservation pipeline \\
        (human processing)\\
        
    };

    \node [block, right of=preservation] (scanning) {
        \includegraphics[width=4.9cm]{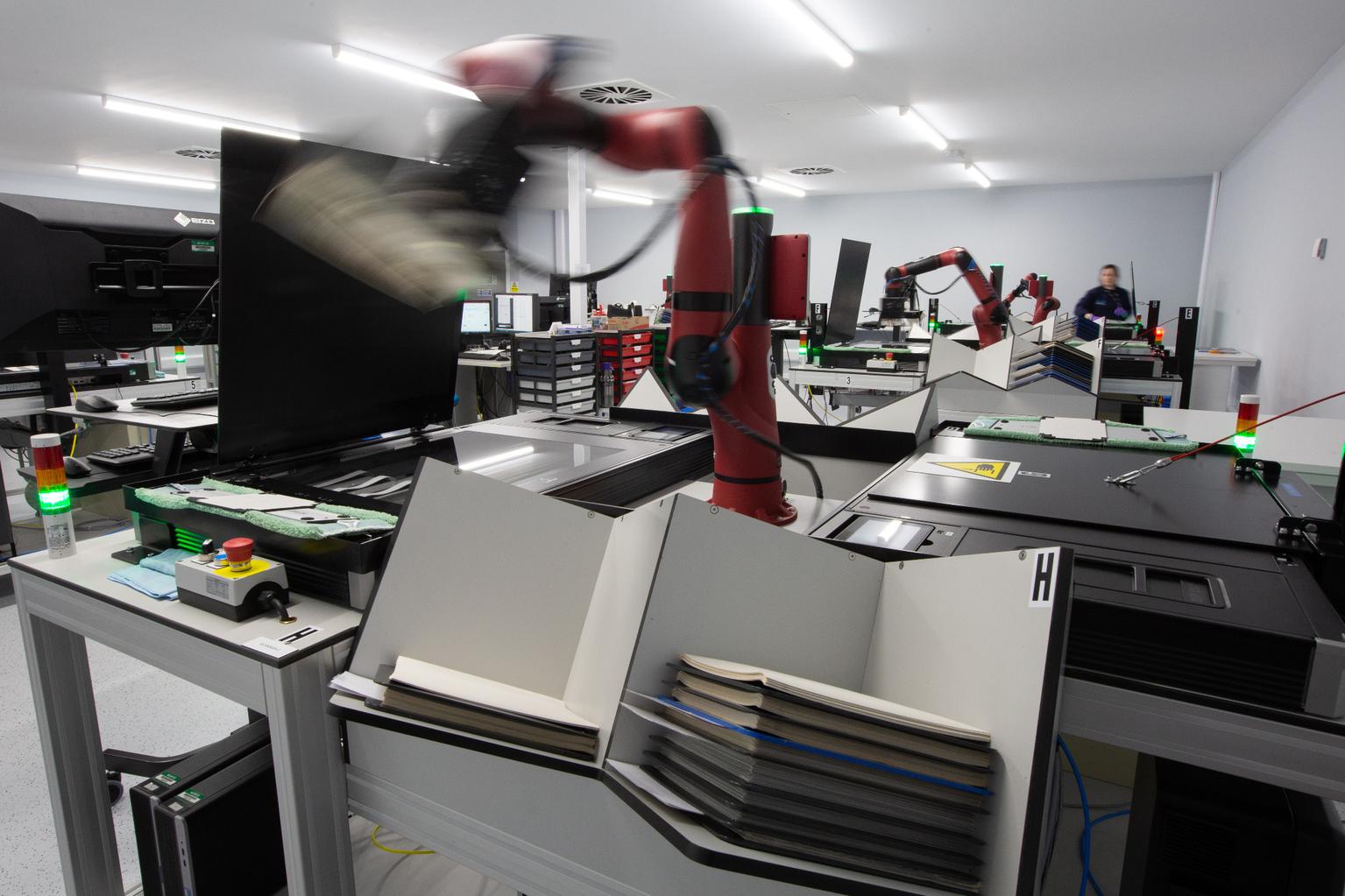}
        \vspace{1ex}\\
        Scanning pipeline \\
        (robotized)
    };

    \node [block, right of=scanning] (postprocessing) {
        \includegraphics[width=4.9cm]{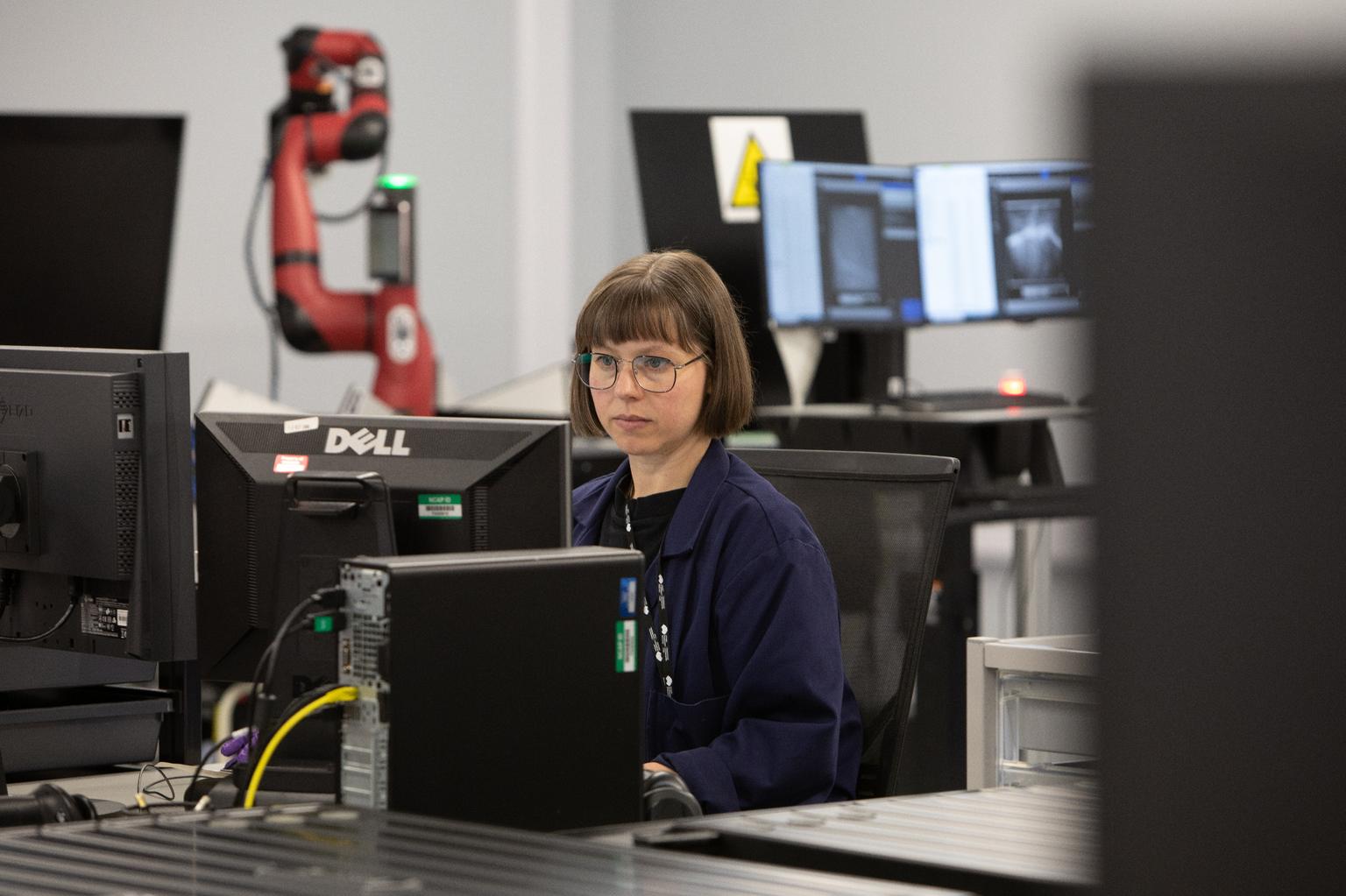}
        \vspace{1ex}\\
        Post-processing \\
        (digital and human)
    };

    % Lines
    \path [line,draw=NavyBlue] (storage) -- (preservation);
    \path [line,draw=NavyBlue] (preservation) -- (scanning);
    \path [line, draw=NavyBlue] (scanning.south) -- ++(0,-1cm) -- ++(-12cm,0) -- (storage.south);

    \path [line,draw=ForestGreen] (scanning) -- (postprocessing);

    % Key
    \node [below of=postprocessing, node distance=4.75cm, anchor=south] (key) {};
    
    \node [right=-1.2cm of key] (key1) {
         \begin{tikzpicture}[baseline=-0.5ex]
            \draw[ultra thick, ForestGreen, ->, >=latex] (0,0) -- (1,0);
        \end{tikzpicture} Digital scans
    };
    
    \node [left=0.7cm of key1] (key2) {
         \begin{tikzpicture}[baseline=-0.5ex]
            \draw[ultra thick, NavyBlue, ->, >=latex] (0,0) -- (1,0);
        \end{tikzpicture} Photographic prints
       
    };

\end{tikzpicture}
}

    \captionsetup{margin={-1cm,-1cm}}
     \caption{\raggedright \textbf{Overview of robot-assisted pipeline} LtR: Prints in storage cartons in temperature and humidity-controlled in-house storage, after retrieval from long-term storage. Inspecting prints to evaluate condition and necessary remediation. Robotic arm lifting new print onto scanner bed. Quality control checks during post-processing. \textcopyright National Collection of Aerial Photography.     
     \label{fig:overview}}
  
\end{figure}

\end{landscape}
}

\afterpage{
\begin{figure}
\thisfloatpagestyle{empty}
\begin{center}
\begin{tikzpicture}

% Nodes in the left column
\node (checkMould) [decision] { \includegraphics[width=4.98cm]{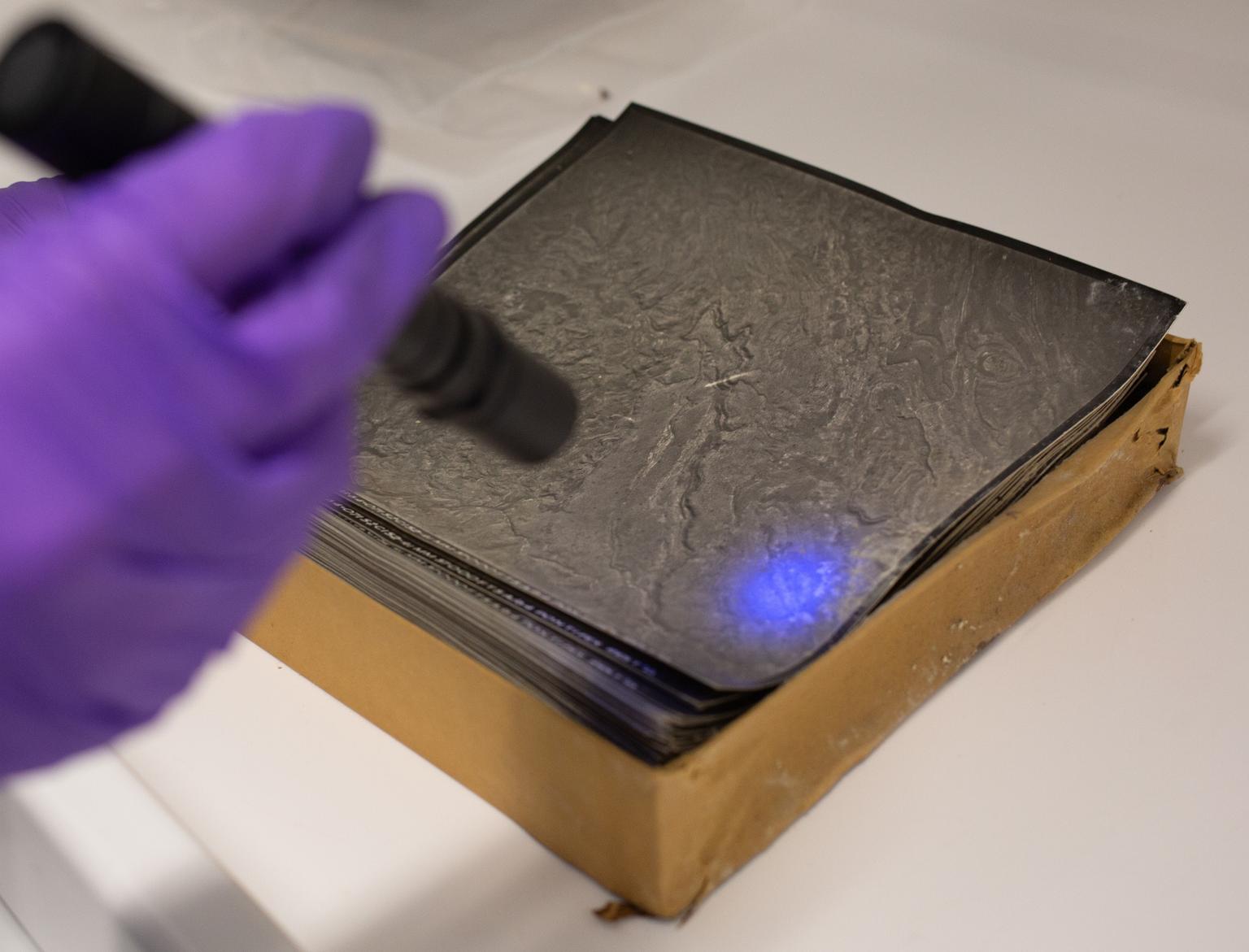}  Evidence of mould?};
\node (checkBlocking) [decision, below=1.5cm of checkMould.south] {\includegraphics[width=4.98cm]{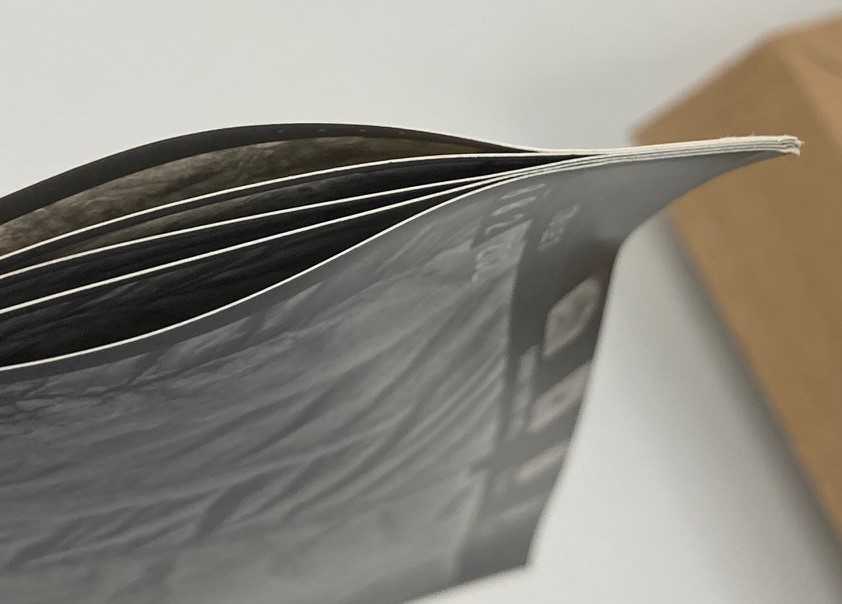} Blocking (adhered prints)?};
\node (int2) [coordinate, below=1.2cm of checkBlocking.south] {};
\node (checkSilver) [decision, below=0.5cm of int2, anchor=north] { \includegraphics[width=4.98cm]{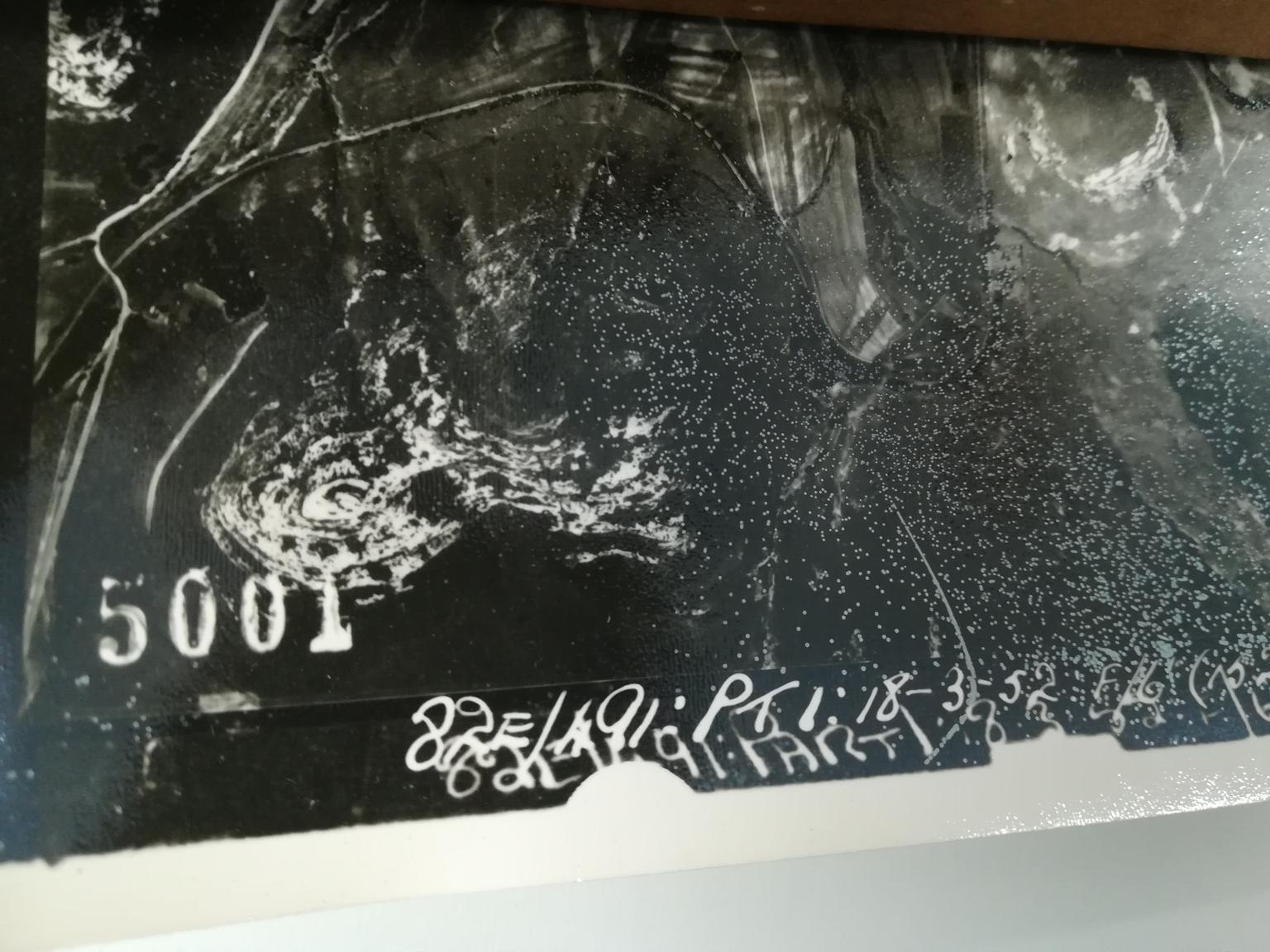} Silver dust?};
\node (int3) [coordinate, below=1.2cm of checkSilver.south] {};
\node (nextPage) [coordinate, below=0.5cm of int3]{};
\node (continued) [below=0.25cm of nextPage] {\emph{continued on next page}};

%nodes in the right column
\node (actionMould) [process, right=1.5cm of checkMould.north east, anchor=north west] {\includegraphics[width=4.98cm]{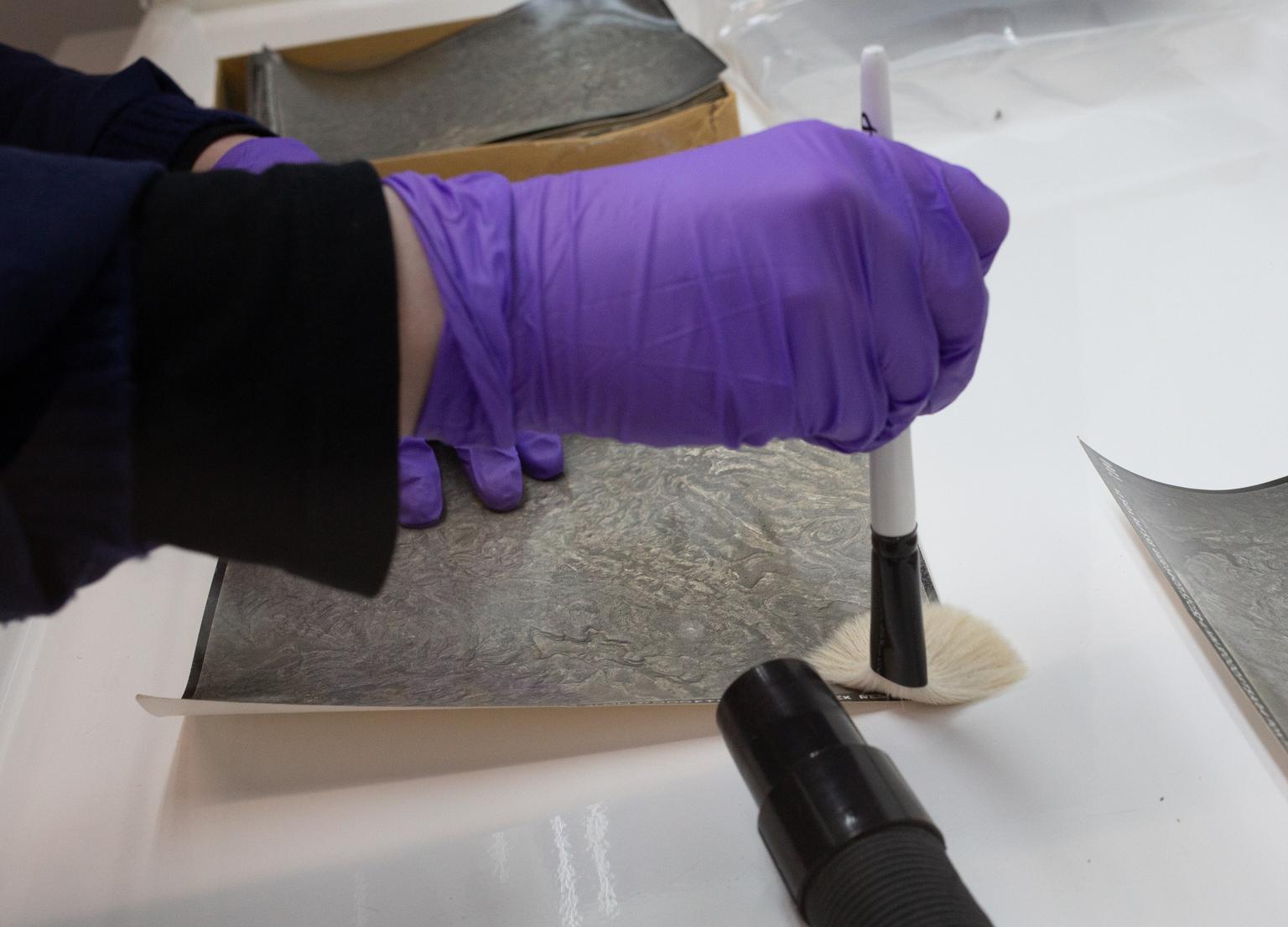} Clean mould; process in separate pipeline.};
\node (actionBlocking)  [process, right=1.5cm of checkBlocking.north east, anchor=north west] {\includegraphics[width=4.98cm]{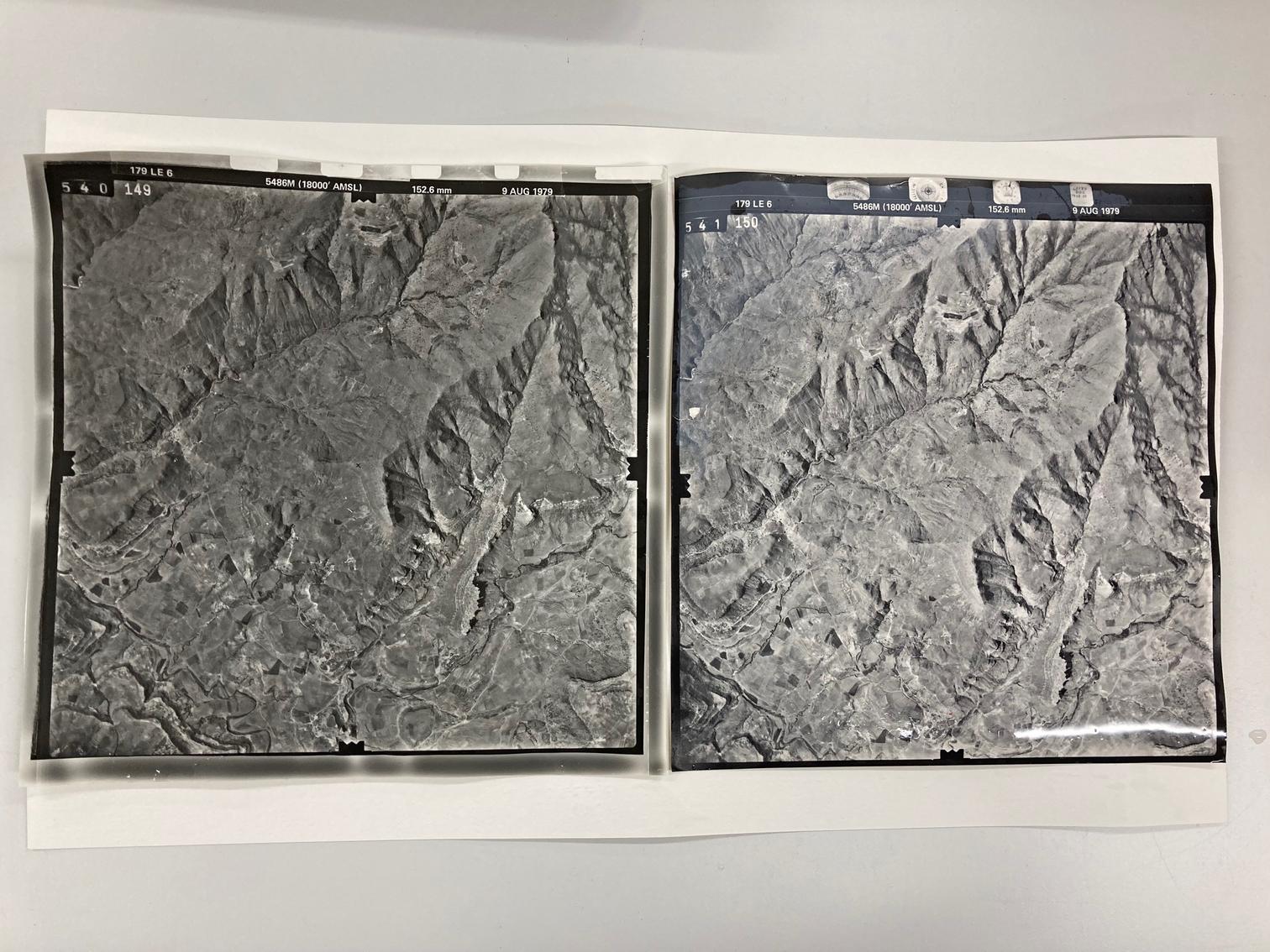} Physically separate, soaking if needed.};
\node (actionSilver) [process, right=1.5cm of checkSilver.north east, anchor=north west] {\includegraphics[width=4.98cm]{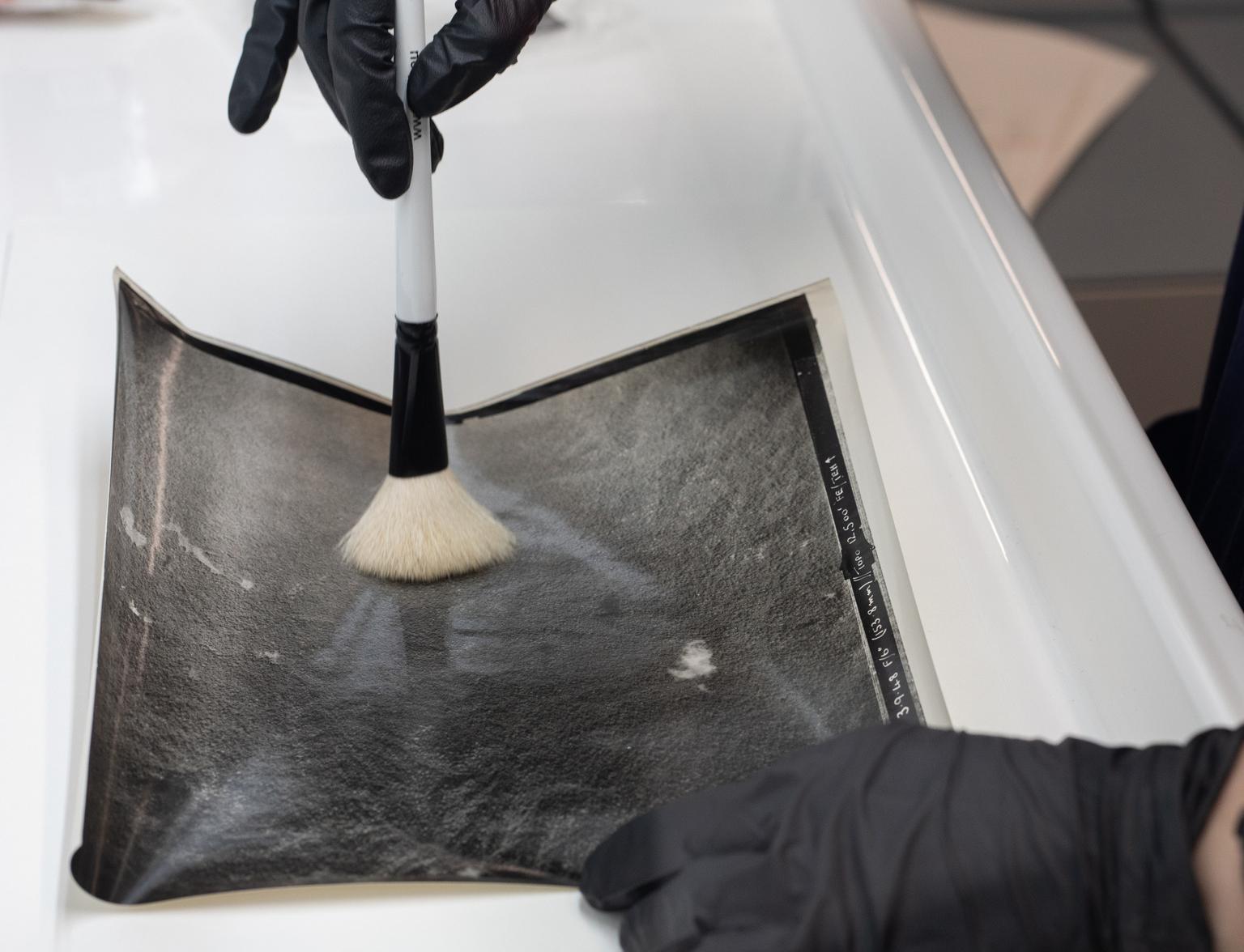} Dry clean with brush or PEC pad.};

% Horizontal "yes" decision paths
%\draw [arrow] (checkMould) -- node[anchor=south, yshift=0.25cm] {Yes} (actionMould);
\draw [arrow] (checkMould.east) -- ++(1.5cm, 0) node[midway, above, yshift=0.25cm] {Yes};
\draw [arrow] (checkBlocking.east) -- ++(1.5cm, 0) node[midway, above, yshift=0.25cm] {Yes};
\draw [arrow] (checkSilver.east) -- ++(1.5cm, 0) node[midway, above, yshift=0.25cm] {Yes};

% Vertical "No" decision paths
\draw [arrow] (checkMould.south)  -- (checkBlocking.north) node[midway, left] {No};
\draw [line] (checkBlocking.south) -- (int2) node[midway, left] {No};
\draw [arrow] (int2) -- (checkSilver.north);
\draw [line] (checkSilver.south) -- (int3) node[midway, left] {No};
\draw [arrow] (int3) -- (nextPage);

% Connecting lines from actions to the intermediate nodes
\draw [line] (actionBlocking.south) -- ++(0,-0.5) |- (int2);
\draw [line] (actionSilver.south) -- ++(0,-0.5) |- (int3);

\end{tikzpicture}
\end{center}

\caption{\raggedright \textbf{Overview of preservation pipeline} TtB and LtR: Checking for active mould using UV light. Removing surface mould using museum vacuum with HEPA filter and brush. Blocked/adhered prints.  Separated blocked prints. Silver dust on surface of print.  Dry cleaning silver dust with brush using a museum vacuum with HEPA filter to collect particulate matter. \textcopyright National Collection of Aerial Photography.\label{fig:preservation}}

\end{figure}
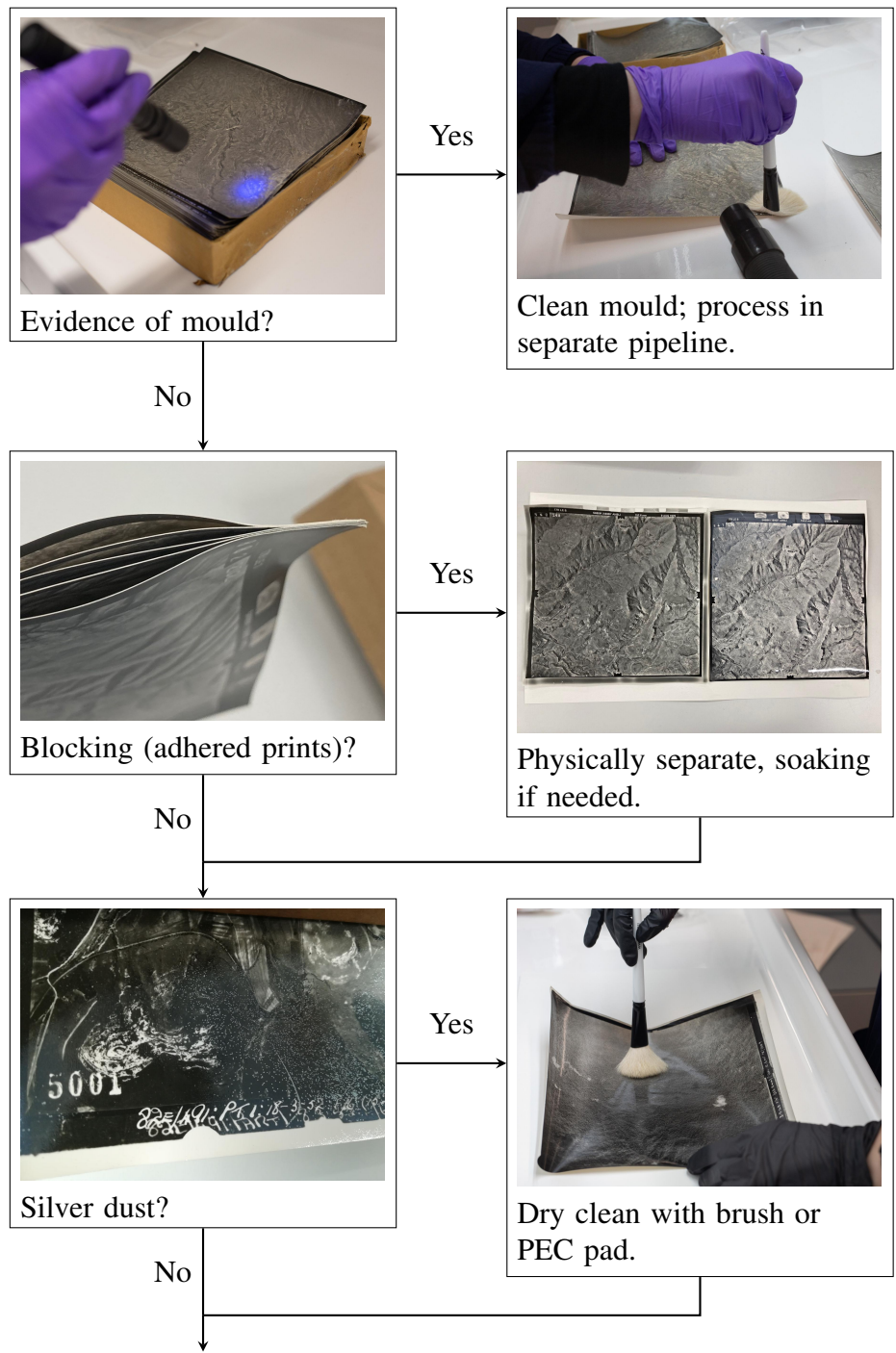

\newpage

\begin{figure}\ContinuedFloat
\thisfloatpagestyle{empty}
\begin{tikzpicture}
% Nodes in the left column

\node (checkAnnotations) [decision] {\includegraphics[width=4.98cm]{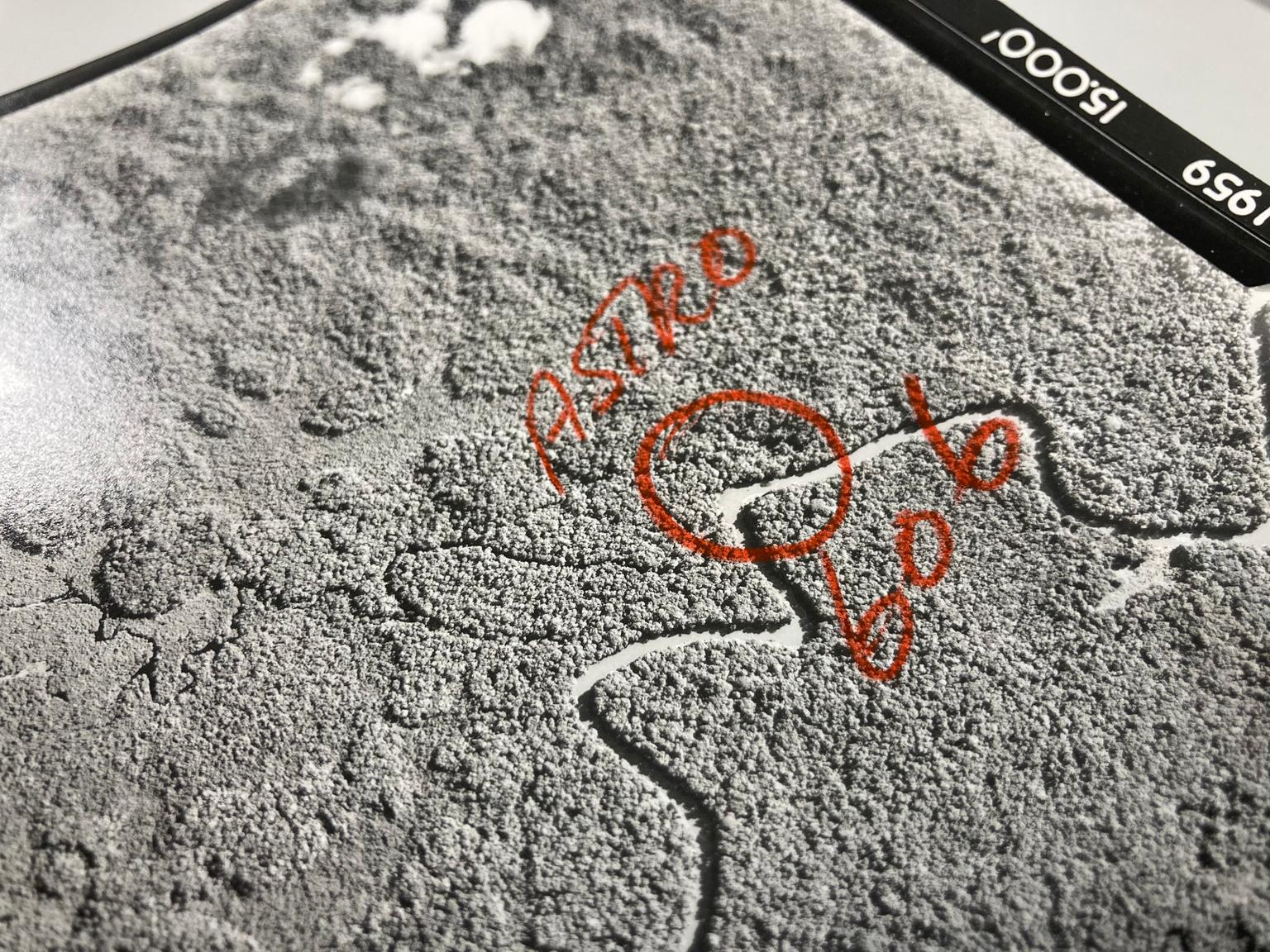} Annotations or adhesives?};
\node (int4) [coordinate, below=1cm of checkAnnotations.south] {};
\node (checkCurling) [decision, below=0.5cm of int4] {\includegraphics[width=4.98cm]{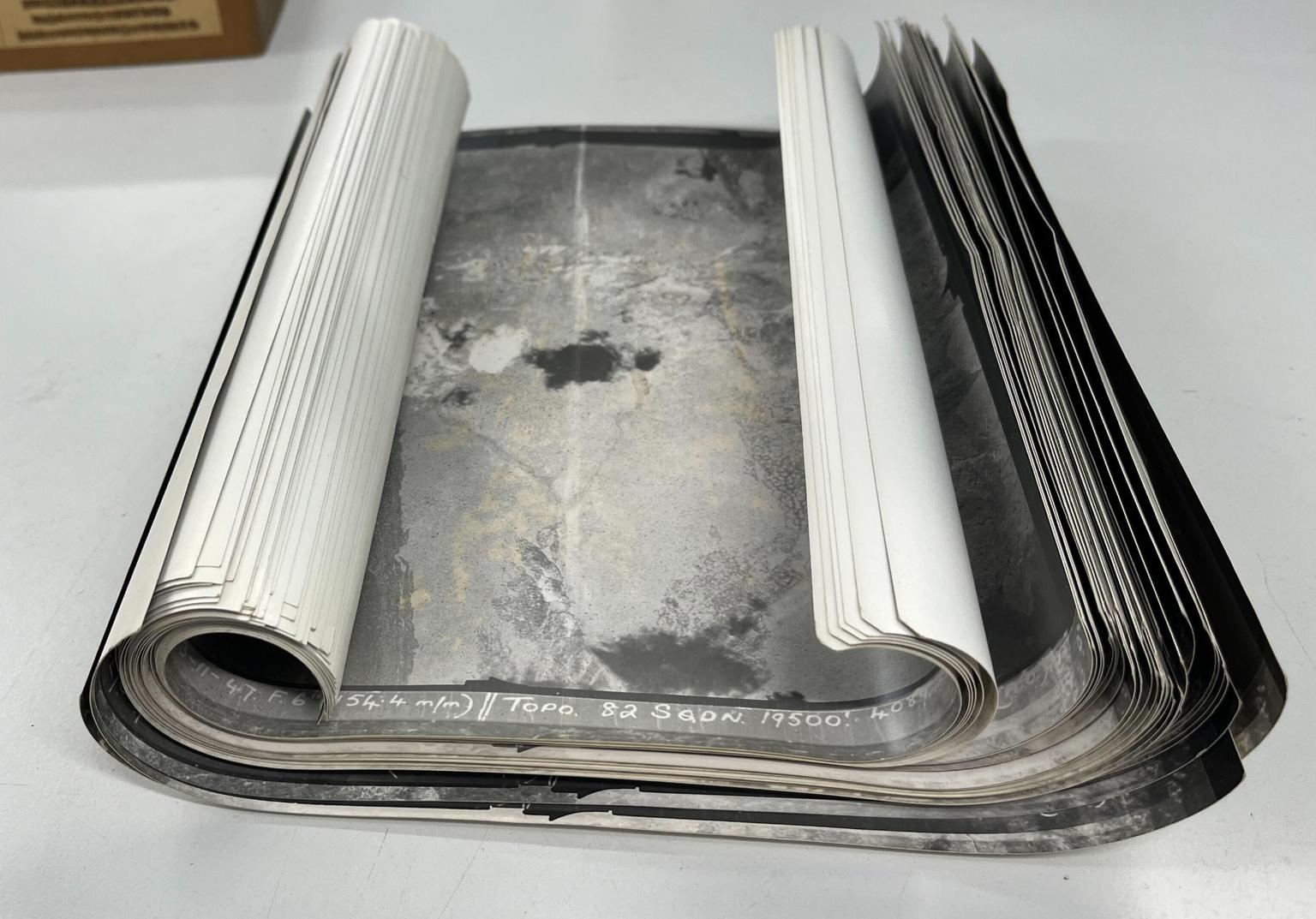} Substantial curling or creases?};
\node (int5) [coordinate, below=1cm of checkCurling.south] {};
\node (checkRips) [decision,below=0.5cm of int5] {\includegraphics[width=4.98cm]{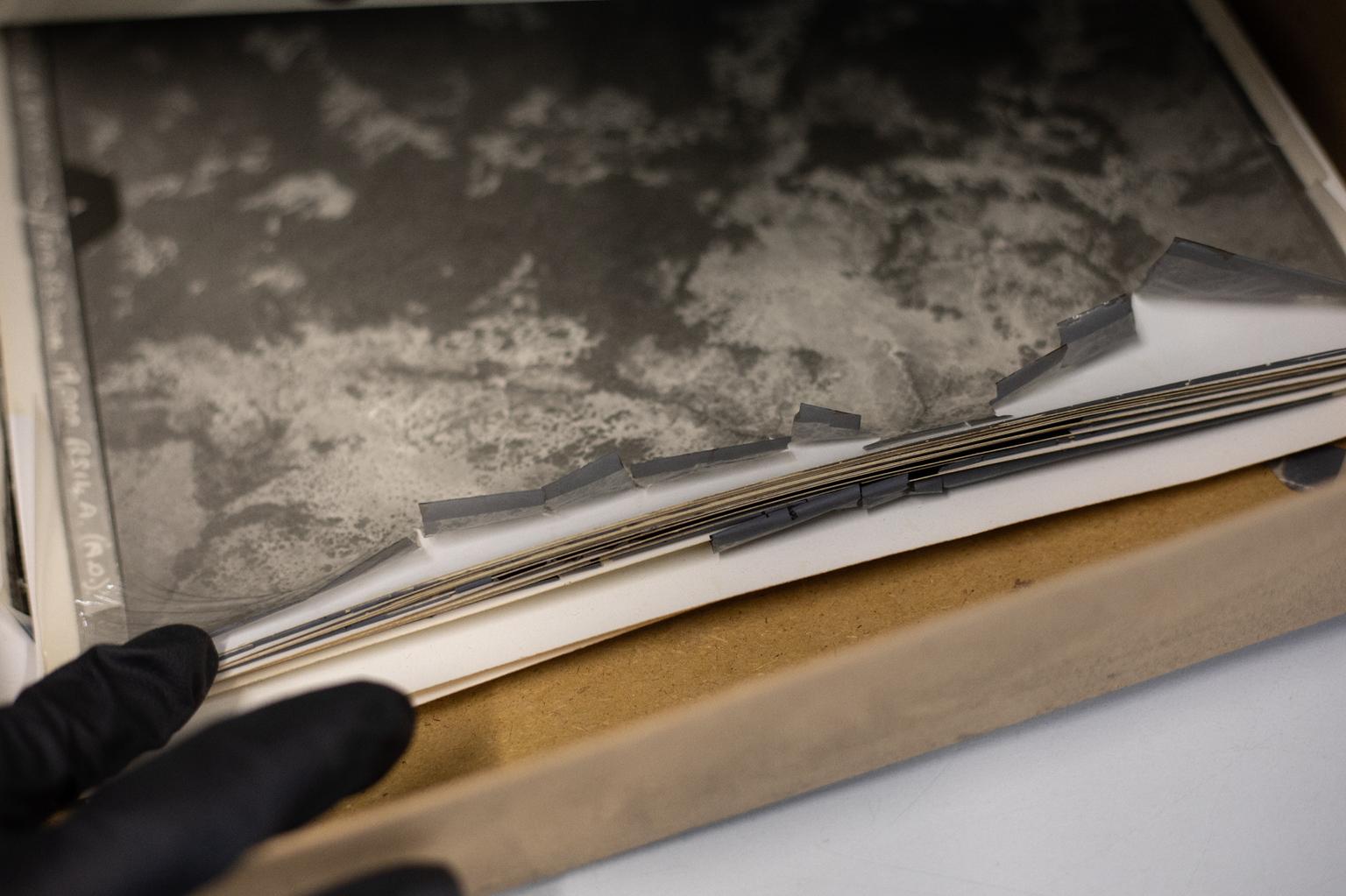} Rips or peeling emulsion?};

%nodes in the right column
\node (actionAnnotations) [process, right=1.5cm of checkAnnotations.north east, anchor=north west] {\includegraphics[width=4.98cm]{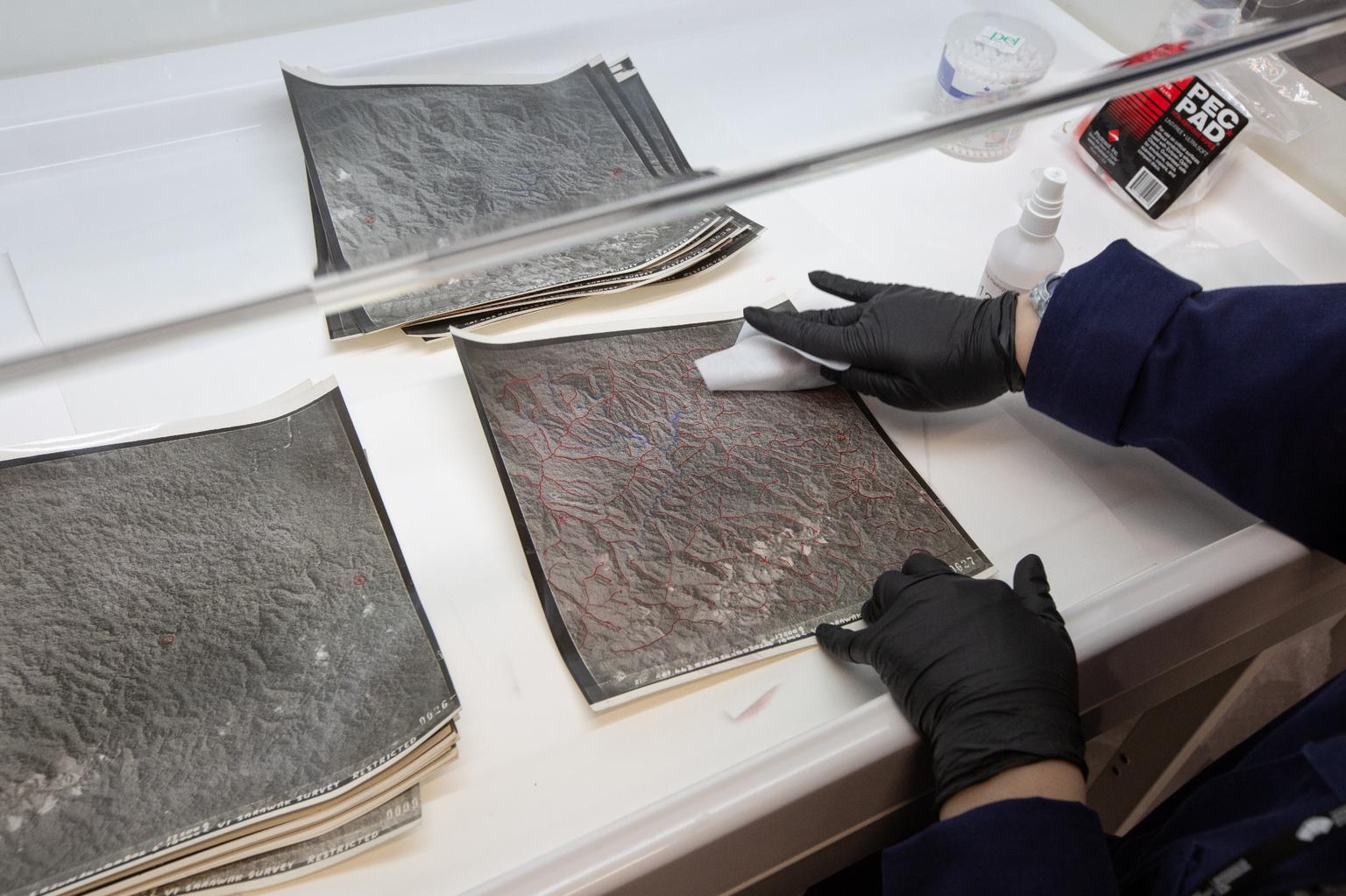} Clean with PEC-12 solvent and PEC Pads.};
\node (actionCurling) [process, right=1.5cm of checkCurling.north east, anchor=north west] {\includegraphics[width=4.98cm]{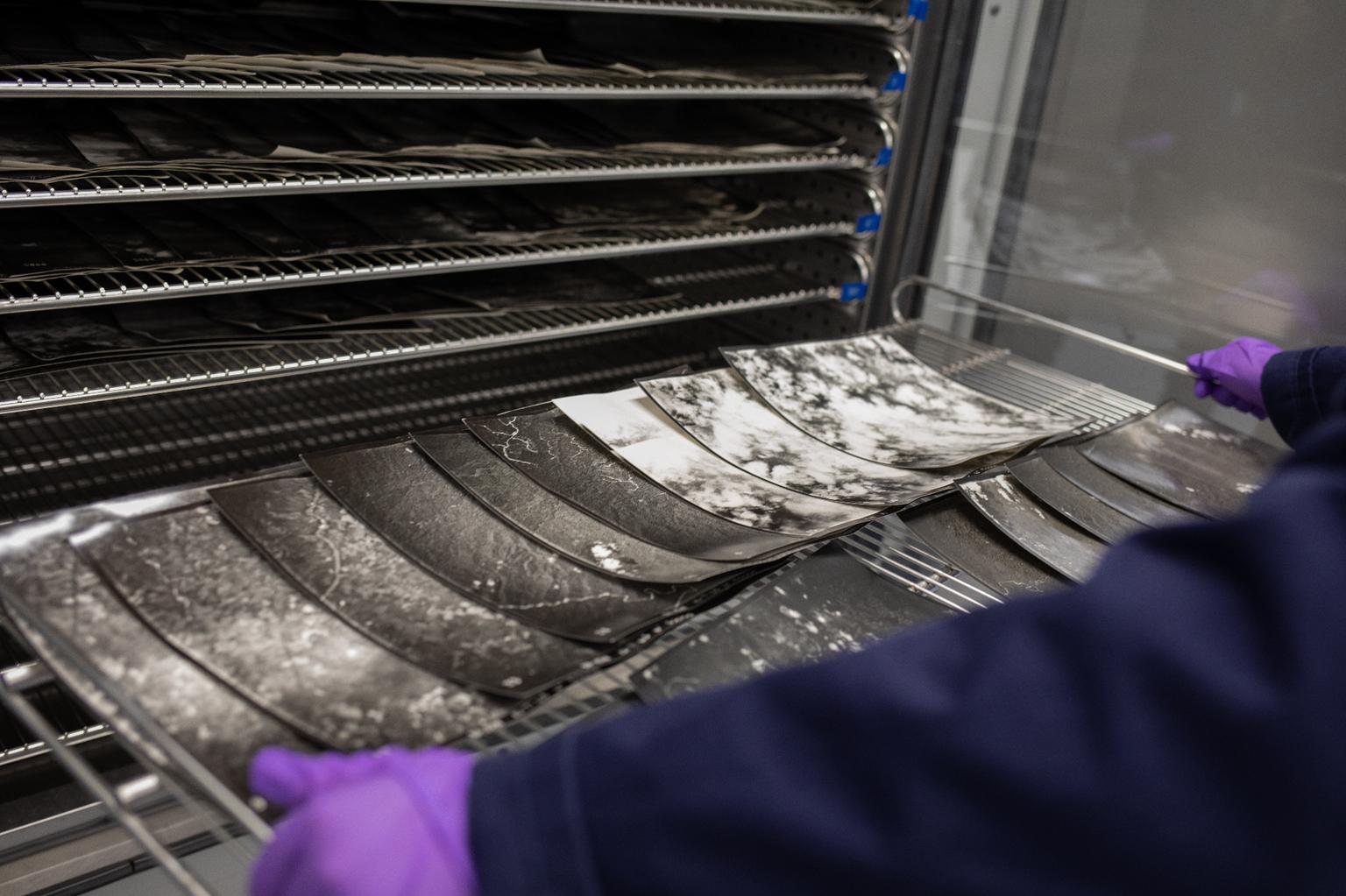} Humidify, then flatten in nipping press.};
\node (actionRips) [process, right=1.5cm of checkRips.north east, anchor=north west] {\includegraphics[width=4.98cm]{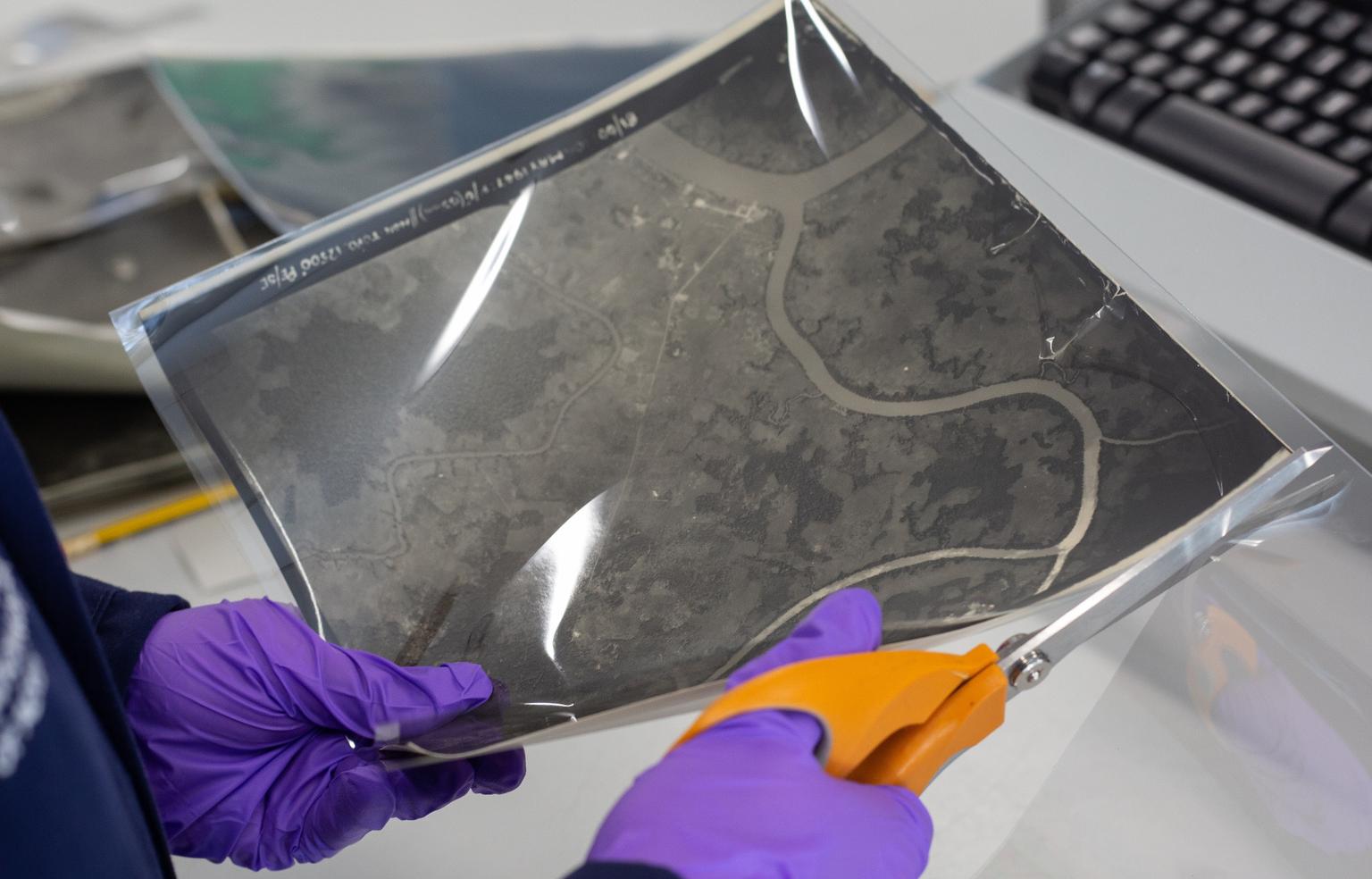} If not too extensive for scanning, protect using polyester sleeves.};
\node (actionEnd) [process, anchor=north, yshift=-1cm] at (actionRips.south) {\includegraphics[width=4.98cm]{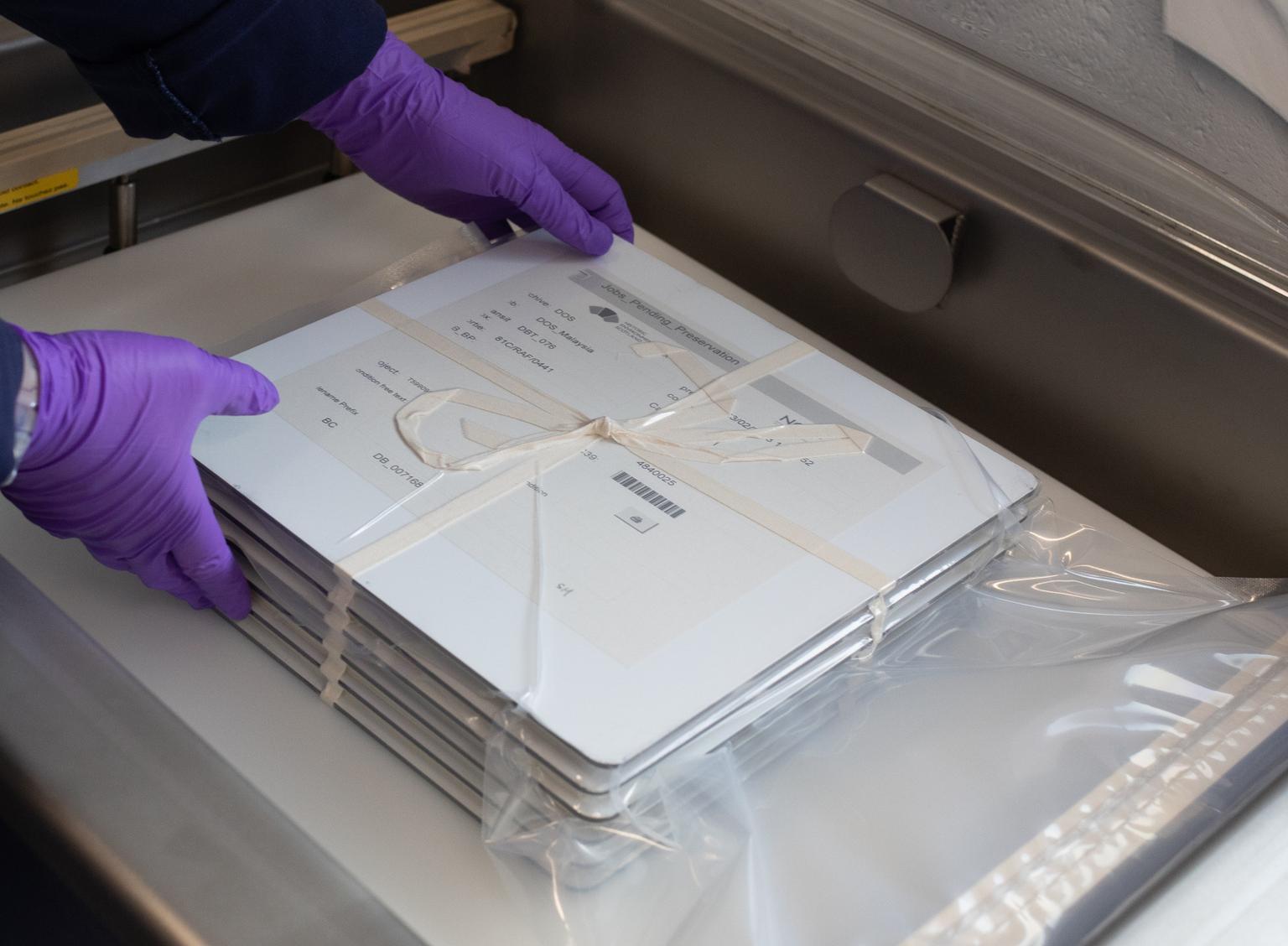} Vacuum pack and pass to scanning pipeline};

% Horizontal "yes" decision paths
\draw [arrow] (checkAnnotations.east) -- ++(1.5cm, 0) node[midway, above, yshift=0.25cm] {Yes};
\draw [arrow] (checkCurling.east) -- ++(1.5cm, 0) node[midway, above, yshift=0.25cm] {Yes};
\draw [arrow] (checkRips.east) -- ++(1.5cm, 0) node[midway, above, yshift=0.25cm] {Yes};
\draw [arrow] (actionRips) -- (actionEnd);

% Vertical "No" decision paths
\draw [line] (checkAnnotations.south) -- (int4) node[midway, left] {No};
\draw [arrow] (int4) -- (checkCurling.north);
\draw [line] (checkCurling.south) -- (int5) node[midway, left] {No};
\draw [arrow] (int5) -- (checkRips.north);

% Connecting lines from actions to the intermediate nodes
\draw [line] (actionAnnotations.south) -- ++(0,-0.5) |- (int4);
\draw [line] (actionCurling.south) -- ++(0,-0.5) |- (int5);

% Final "No" arrow from checkRips to actionEnd
\draw [arrow] (checkRips.south) -- ++(0,-0.5cm) node[left, xshift=-0.5cm] {No} -- ++(0,-2.5cm) -- ++(4cm,0);

%pseudo caption
\node (caption) [left=0.2cm of actionEnd, yshift=-1.8cm, text width=10cm, align=left, inner sep=0] {\raggedright \textbf{Overview of preservation pipeline, cont.} TtB and LtR: Chinagraph annotations. Cleaning using PEC-12 solvent and PEC Pads. Curling prints. Preparing prints for the humidification chamber before flattening. Print with extensive emulsion peeling. Protecting a print with emulsion peeling using a polyester sleeve. Vacuum packing stacked prints, interleaved with steel plates, ready for scanning. \textcopyright National Collection of Aerial Photography.};

\end{tikzpicture}
\end{figure}

}

\subsubsection*{The manual preservation pipeline} 

Before prints can be scanned, workers need to evaluate the condition of each photograph, undertake remedial work where necessary, and prepare the images for scanning (Fig. \ref{fig:preservation}). The prints are silver gelatin, made up of several layers: a paper backing; a white pigmented layer; the photosensitive layer, a suspension of light-sensitive silver salts in gelatin, referred to (albeit erroneously) as the emulsion; and a protective coating \cite{lavedrine2009photographs,stulik2013atlas}. Due to inadequate storage conditions in the past, some prints have sustained damage. Evaluating and remediating damage requires extensive skill and discretion. The range of preservation issues that can occur is substantial. Table \ref{tab:preservation} summarizes the total number of boxes and the share affected by specific preservation issues. The most commonly-encountered issues are as follows:

\paragraph{Mould} When stored in conditions that are too humid---as they were prior to NCAP's custody---mould spores can infest prints, feeding off the gelatine and sometimes the paper itself\cite{lavedrine2009photographs}, leaving spores or stains. We remove surface mould using using a museum vacuum with a HEPA filter and a brush. Much of the mould is dormant, but we evaluate activity by checking for fluorescence with an ultraviolet torch. If mould is active, we may aqueous clean the prints with a solvent and isolate the affected prints in conservation grade polyester sleeves. If the original box is mouldy, we discard it---noting that information from notations on the box has previously been stored photographically---and transfer the prints to a new archival quality box. Evidence of mould necessitates an entirely separate processing pipeline, avoiding all contact with moisture, which could reactivate dormant mould. Boxes of photographs that are affected by mould must also be scanned separately, and any contact surfaces disinfected afterwards, to avoid cross-contamination of mould strains across boxes of prints. 

\paragraph{Blocking} Blocking occurs when two or more images are stuck together. Blocking typically results from exposure to humidity: the gelatine layer absorbs water, becoming tacky and adherent \cite{lemmen2017blocked}. 
Blocked prints can sometimes be separated by gently easing them apart at the corner, especially if the gelatine remains some humidity. Prints are harder to separate when the gelatine is dry \cite{lemmen2017blocked}. When physical separation is not initially successful, prints are soaked in filtered water for several hours and then gently separated using a bamboo spatula. These prints must then be dried in a nipping press over several days, with repeated changes of blotting paper.

\paragraph{Cleaning} Some prints were annotated, typically to highlight visual points of significance such as roads, rivers, vegetation, or other landmarks that a surveyor might need to use as a control point. However, these annotations obscure the features on the image. Annotations were usually made using a grease, or chinagraph, pencil, and can be straightforwardly removed by aqueous surface cleaning with PEC-12 solvent and PEC Pads. Other prints are also affected by silver migration, which occurs as the emulsion breaks down, allowing silver to migrate through the paper to its surface \cite{lavedrine2009photographs}. This results in a sparkly dust that is highly visible on the fingertip of a black nitrile glove and can include masses up to 1mm in length. This dust can be gently brushed or wiped off with a brush or PEC pad, using a museum vacuum with HEPA filter to collect particulate matter.

\paragraph{Curling}  The layers of the photographic paper can expand or contract at different rates when exposed to different humidity levels, resulting in a tendency to curl \cite{swan1981conservation}. Curled prints are vulnerable to damage during robotic handling and scanning. They can fold and crack or crease under the weight of the scanner lid or the steel plates. Additionally, curled areas that do not lie flat against the scanner glass may fall outside the scanner's focal depth of field. Creasing can also arise if images have not been stored flat. Curled or creased images cannot simply be mechanically flattened without risking cracks in the emulsion. Prints with substantial curling or creasing are thus first humidified for 2-3 hours in a large humidifying chamber. After humidification, prints can be pressed flat in a cast iron nipping press between sheets of aluminium composite material (ACM) and wooden boards, with alternating layers of blotting paper, wax paper, or bondina---a non-woven polyester material---to absorb moisture. The drying and pressing process takes several days. Prints that are both curled and affected by mould are humidified and flattened separately without direct contact with water, cleaning or disposing of any materials with which they come into contact.

\paragraph{Rips or peeling emulsion}  Some prints have physical damage, including peeling of the emulsion away from the backing paper \cite{neblette1977neblette}. Prints with significant rips or peeling emulsion are placed in protective polyester sleeves for transport and scanning. Rarely, sorties have such substantial emulsion peeling that they cannot be scanned as so little information is left on the prints. The affected prints are typically from the older sorties, mostly flown by the RAF. In some of these cases, NCAP held the original film in its archive, which could be scanned in lieu of the prints. \\

\noindent The output of the preservation pipeline is a stack of prints ready to scan, vacuum-packed for safe temporary storage while awaiting scanning. 

\subsubsection*{The fully-automated scanning pipeline} 

In contrast to the preservation pipeline, scanning involves a very large number of identical repeated tasks. Beyond loading and unloading the prints, the scanning pipeline is entirely automated. We developed the scanning pipeline through extensive piloting and refinement, including a full pilot trial digitizing 8,000 images from The Gambia. 

Each scanning station houses a multipurpose robotic arm with a bespoke end-of-arm tool designed to lift and release images.  The robotic arm is a Sawyer and Sawyer Black robotic arm\cite{rethinkrobotics2019sawyer}, adapted for our task by CBM-Logix, an automation engineering firm specializing in robotics and system integration. Each robot arm houses an optical sensor, four small electromagnets, and a vacuum attachment with four suction cups. The vacuum attachment is calibrated to lift exactly one image from the stack on the input hopper. Each robotic arm operates a pair of Contex IQ Flex scanners (Fig. \ref{fig:scanning}). 

Workers place prints face down in the input hopper, correctly oriented so that each print is placed consistently on the scan bed. A printout summary of the sortie and frame information is placed on top of each stack of prints from a box. Prints are interleaved with matte black or blue powder-coated mild steel plates to maintain flatness of prints while stacked and to ensure stack stability. To further reduce instability, workers only stack prints of the same size. Approximately 300 prints can be loaded at one time. Workers replenish hoppers  throughout the day. After loading prints into the input hopper, workers initiate the robotic programme. 
 
Upon initiation of the robotic programme, the robot arm moves to the input hopper and its optical system detects whether the target is the reverse side of a print (light in colour) or a steel weight (dark in colour).

\afterpage{
\begin{landscape}
\begin{figure}[p!]
\thisfloatpagestyle{empty}

\centerline{
\begin{tikzpicture}

% Define the height of the figure
\def\figureheight{12cm}
% Define the width of the text box
\def\textboxwidth{6cm}

% Include the image on the left
\node[anchor=south west, inner sep=0] (image) at (0,0) {\includegraphics[height=\figureheight]{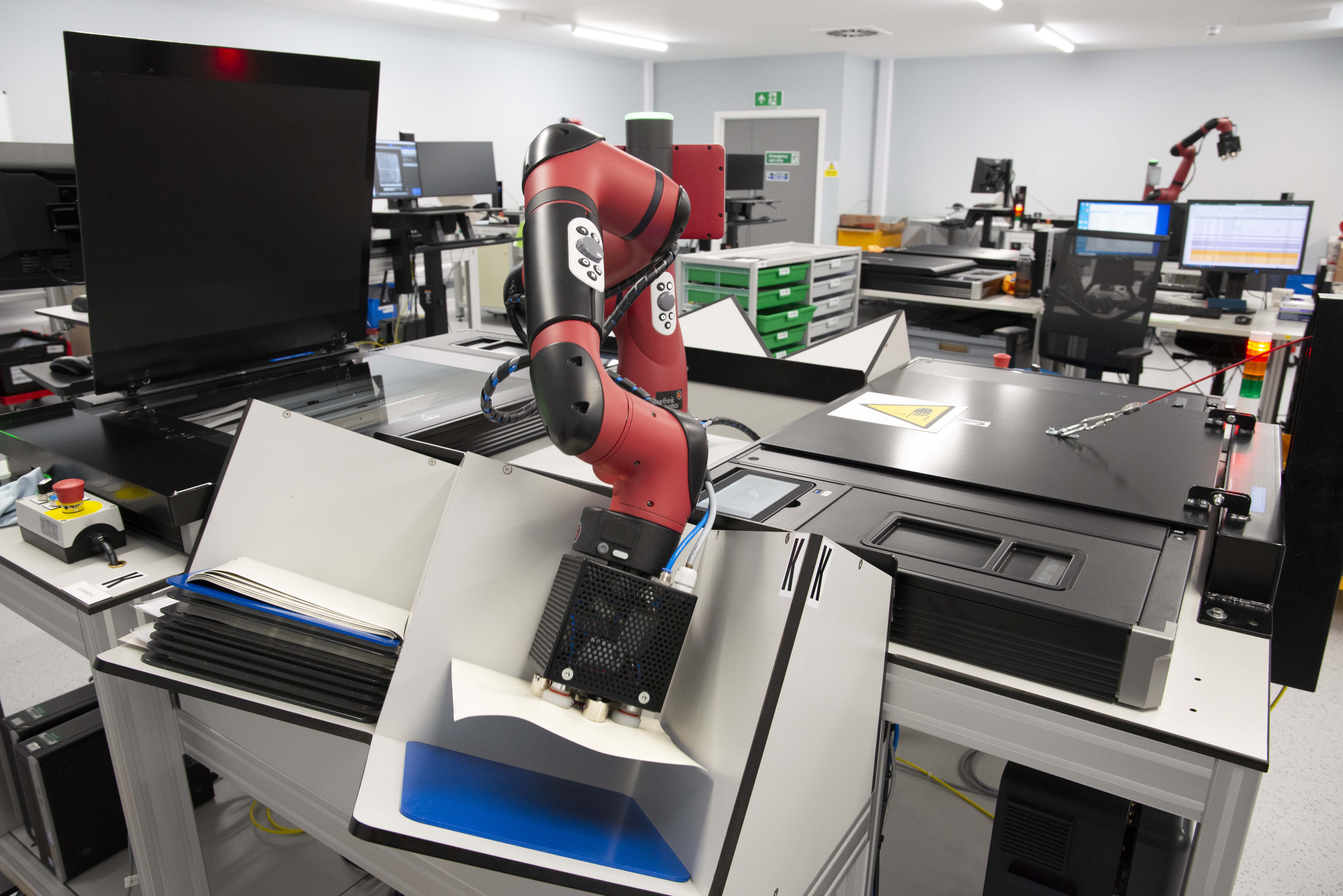}};

% Draw the grid for locating annotations (comment out in final version)
%\draw[step=1cm,gray,very thin] (0,0) grid (18,12);

% Draw the x and y axes for locating annotations  (comment out in final version)
%\draw[thick,->] (0,0) -- (18,0) node[anchor=north west] {x};
%\draw[thick,->] (0,0) -- (0,12) node[anchor=south east] {y};

% Draw numbers in circles on the image
\node[fill=white,draw=black,circle,inner sep=2pt] at (7.5,1) {1};
\node[fill=white,draw=black,circle,inner sep=2pt] at (7.5,9.75) {2};
\node[fill=white,draw=black,circle,inner sep=2pt] at (8.5,3.5) {3};
\node[fill=white,draw=black,circle,inner sep=2pt] at (16,6.25) {4};
\node[fill=white,draw=black,circle,inner sep=2pt] at (3,9) {5};
\node[fill=white,draw=black,circle,inner sep=2pt] at (5.5,7) {6};
\node[fill=white,draw=black,circle,inner sep=2pt] at (13.5,6.5) {7};
\node[fill=white,draw=black,circle,inner sep=2pt] at (4.5,4.5) {8};
%\node[fill=white,draw=black,circle,inner sep=2pt] at (17,8.75) {9};
\node[fill=white,draw=black,circle,inner sep=2pt] at (10.5,7.75) {9};
\node[fill=white,draw=black,circle,inner sep=2pt] at (2,5) {10};
\node[fill=white,draw=black,circle,inner sep=2pt] at (17.5,7.5) {11};

% Add key to the right of the image
\node[anchor=north west, align=left, text width=\textboxwidth] at (image.north east) {
\vspace{-\baselineskip}
\begin{enumerate}[label=\protect\circled{\arabic*}, left=0pt, nosep]
    \item Input hopper, with metal plate.
   \item Robot arm, positioned to unload and reload one scanner while a second scans.
   \item Bespoke end-of-arm tool comprising an optical sensor to distinguish light prints and dark plates; four small electromagnets to lift plates; and a vacuum attachment with four suction cups to lift prints. 
   \item Automated pulley system for scanner lid. 
   \item Scanner lid, lifted, allowing robot arm access to place or remove print. 
   \item Scanner bed, for print placement.
   \item Scanner lid, closed during scanning.
   \item Output hopper, storing scanned prints interleaved with mild steel plates to keep the stacked prints flat and stable.
   \item Second input and output hoppers.
   \item Control panel and emergency stop. 
   \item Status light. 
\end{enumerate}
\vspace{-\baselineskip}
};

\end{tikzpicture}

}     
     
     \captionsetup{margin={-2.5cm,-2.5cm}}
     
     \caption{\raggedright \textbf{Robotic arm and scanner system} Figure shows system as robotic arm (3) lifts a print from the input hopper (1) to load it onto the scanner bed (6). The scanner lid (5) will close automatically and the scan initiates. While this scan completes, the second scanner lid (7) lifts and the robot arm will load a print from the second scanner's input hopper (9) onto the second scanner bed. Once the first scan is complete, the scanner lid lifts, and the robot arm lifts the print from the scanner bed to the output hopper (8). \textcopyright National Collection of Aerial Photography. \label{fig:scanning}}
\end{figure}
\end{landscape}
}

If a print is detected, the arm initiates the vacuum and contacts the print until the pressure changes, indicating good contact. If no pressure change occurs, the robot arm makes two further attempts, applying a small additional downward force before initiating the vacuum. This force ensures that any curled prints will be flattened and in good contact with all four suction cups. The robot arm then lifts the print from the hopper and moves it to the scanner bed, slowly lowering the print onto the bed until it detects a predetermined level of resistance indicating contact with the scanner bed. The vacuum is turned off and the arm moves clear. If a weight is detected, the arm aligns centrally with the weight and lifts it using the electromagnets.  The robot arm then transfers the weight to the output hopper. If unsuccessful, the robot arm makes two further attempts before stopping and reporting an error. 
 
The scanner lid is automated using a digitally controlled modified gate opener to cycle the lid. Once the robot arm is clear, it signals to the pulley system to lower the scanner lid.  Once the lid is lowered, the lifting system in turn signals to the scanner to initiate a scan. 

Once a successful scan is confirmed, the process is reversed. The successful scan triggers lifting of the lid.  After the lifting cycle is complete, the controller signals to the robot arm to lift the scanned print from the scanner bed and move it to the output hopper, and then to reload the scanner. While each scan is underway, the robotic arm rotates $180^{\circ}$ to unload and reload a second scanner. After the last image is removed from the input hopper, a small lamp lights up. When the robot arm next attempts to pick a print from the input hopper, the optical sensor detects the light, and production stops on that scanner awaiting reloading. The robot arm continues to load the other scanner. 

All loading cycle transitions are triggered using absolute references, such as successful lifting of a print or completion of a scan. This eliminates the need for error margins, maximizes efficiency, and ensures that movement cycles only initialize when it is safe.

The scanning pipeline operates in a controlled air environment, maintained at a weak positive pressure through a system of airlocks. Humidity and temperature control are vital when working with archival prints. Additionally, prints shed fibers which can contaminate scanning work spaces, creating tiny artefacts on scanned images.  In operation, the controlled air environment successfully reduces fiber spread and greatly delays the onset of print curl.

The scanners are integrated with Nextimage software and set to scan at  1200 ppi, the highest optical resolution of the scanners. The choice of 1200 ppi is partly driven by practical considerations, including the limited availability of scanners that undertake higher-resolution scanning at speed, but it also compares quite favourably to optimal scanning resolutions derived from sampling theory \cite{light1993national}. For an print with resolution $R$ defined in line pairs per millimeter, the optimal scanning pixel size lies between $\sfrac{1}{2\sqrt{2}R}$ and  $\sfrac{1}{2R}$. For 27 line pairs per millimeter, the acceptable pixel range is between 13 and 18~$\mu$m; for 10 line pairs per millimeter, the acceptable range is between 35 and 50~$\mu$m. At 1200 ppi resolution, pixels are approximately 21~$\mu$m, more than sufficient for 10 line pairs per millimeter and slightly higher than ideal for 27 line pairs per millimeter. At 1200 ppi, we thus largely preserve the information in the original photographs. 

The manufacturer reports a scanning accuracy of 0.1\% +/- 1 pixel. Each scanner bed is pre-loaded with a 250 x 25 mm calibration target, displaying a measure scale, tonal step wedge, and resolution target (Fig \ref{fig:calibration}). The pre-set scan area is 340 x 315mm, except for the largest 22 x 11.5 inch prints. The scan area includes a margin around the print and the calibration target (see Fig \ref{fig:scan}). Except for the few colour prints, we scan at 8-bit grayscale,

\subsubsection*{Post-processing, quality control, validation, and distribution} 

Once a sortie is completely scanned, it is registered as complete in the database, and scanning metadata is recorded, including the scanning date and the scanner. Scans are checked and renamed to correspond to the print numbers. Workers identify missing scans or prints that were flagged during preservation as too damaged for robot handling. They then locate the corresponding prints and scan them manually on a dedicated flatbed scanner. An automated procedure crops the NCAP calibration target and the surrounding black background, leaving the print and a 5 mm black border. 

A worker later checks the quality of the scan and the cropping. They identify any folds or dust particles obscuring the image surface, issues with file naming---either at sortie level or frame numbers---and confirm that none of the image has been cropped. Any prints that fail these quality checks are renamed, re-cropped, or recalled for re-scanning. Material is returned to the off-site storage facility after completing the quality checks.

Capturing the calibration target allows us to later verify the size, tonal values of the image, and capture resolution. In our stratified sample of 124 images, we manually labeled the calibration targets to evaluate the consistency of the results. First, we labeled the beginning and end points of the measure scale in inches and calculated the length of the scale in pixels. Second, we labeled the approximate centroids of the first and last bar of the tonal step wedge, interpolated points between them to locate the centroids of the intermediate bars, and recovered the grayscale pixel value for each bar's centroid on each image. Lastly, we inspected the resolution targets to identify the last group of bars that could be seen and correctly counted, which reveals the size of the smallest distinguishable feature, to compare to the theoretical resolution of the scan. 

The primary uncropped archival versions created by the scanner are uncompressed, full-resolution TIFFs, each approximately 250 MB in size. The entire digitized DOS archive thus constitutes around 425 TB. These versions are stored in archival-grade digital storage optimized for long-term preservation and infrequent access. We also create cropped, 1200 dpi, 50\% compressed JPEG versions of each image for distribution to users. Once georeferenced, these versions will be searchable via a mapping interface and available for download on NCAP's website (https://www.ncap.org/). Since NCAP is a public organization that operates on a cost recovery basis, website subscriptions and download fees fund the costs of site maintenance and service. 

\section*{Results}

\subsection*{A thirty-fold increase in worker productivity}

Human workers can, in principle, achieve higher scanning rates per scanner-hour than the automated system. Each scanner can produce a 1200 pixel per inch (ppi) scan in 45 seconds, whether operated by a human worker or a robot. A human worker continuously operating 2 scanners can achieve a scanning rate of 100 scans/hour, or 50 scans/scanner-hour. While the average time for a loading cycle in the automated system varies, as a result of the use of absolute references to trigger transitions, we estimate using several days of hourly production records that a robot operating two scanners can on average achieve scanning rates of 54 scans/hour, or 27 scans/scanner-hour, 54\% of the theoretical maximum human worker rate. 

One human worker, however, can only load and unload two scanners manually and only operates scanners during working hours. Assuming a 35-hour work week gives productivity of 1750 scans/scanner-week and 3500 scans/worker-week. Each human worker can load and troubleshoot four robots and eight scanners in only 2 hours of their working day, leaving the remaining hours free for other tasks, such as quality control. The robotic systems can also, in principle, load and unload scanners 24 hours a day, 7 days a week. Assuming 24-hour production 7 days a week gives weekly productivity of 4536 scans/scanner-week and 127,008 scans/full-time-worker-week. Per scanner, the increase in productivity of the robotic system over the manual system is 2.6-fold. Per full-time-worker-week, the increase in productivity is more than thirty-fold. 

In practice, neither the manual system nor the robotic system will consistently achieve these theoretical maximum production rates. Humans can be distracted or need to take breaks and sick days, and replacement workers need training and time to learn on the job. If the robotic system encounters a problem---a failed lift of a print, for example---it pauses and returns an error, which may not be resolved until the next working day if it happens at night. 

The equipment for this project comprised 7 robots and 14 scanners, giving a theoretical maximum productivity---if all 14 scanners operated continuously for 24 hours a day and 7 days a week---of 9,072 scans/day and 63,504 scans/week. While the maximum 24-hour scanning rate achieved in this project was almost exactly equivalent to this theoretical maximum, the maximum weekly scanning rate in this project was just over half the theoretical maximum. Team members recorded daily and weekly production statistics, annotated to record reasons for anomalous events. Production statistics were only recorded on weekdays, so statistics recorded on Mondays include production over the weekend. Excluding Mondays, the maximum daily scanning rate was 9,090 scans, effectively identical to the theoretical maximum. The maximum weekly scanning rate was 36,084, 56\% of the theoretical maximum. There was, however, significant learning by doing: scanning rates increased by an average of 13\% week-on-week for the first eleven months of this project. After this point, scanning productivity exceeded preservation productivity, creating a bottleneck in the process. 

\subsection*{Safe for workers and protective of archival material}

The pipeline was both safe for workers and protective of the irreplaceable archival material. 

No worker sustained any more than minor injuries during the project. Workers wore gloves, masks, and protective lab coats, and they cleaned prints in fume cabinets to allow the safe removal of mould without exposure to mould spores and to avoid exposure to fumes from solvent cleaners. The Sawyer robotic arm is intrinsically designed for safe use as a collaborative robot, with limited power and force capabilities \cite{rethinkrobotics2019sawyer}. Design features and operational specifications, including shock-absorbing springs in every joint and speed limits on robotic movement, minimize the force associated with any human-robot collision. Additionally, the motion footprint of the robotic arm is designed to minimize the risk of collision. During this project, we recorded one incident of a human-robotic arm collision, which resulted, as designed, in no significant injury. 

 We estimate that more than 99.9\% of prints were scanned without damage. Any damage incurred was minor, limited to print corners or curled prints being caught and folded by the scanner lid or steel plates, and some prints coming into contact with the compressor fluids from the robotic arm. All damage incurred could be remediated using standard preservation techniques and did not affect the fidelity of the final scan. 

\subsection*{Fidelity of reproduction}

The digital images faithfully reproduce the printed images. Inspecting the calibration targets shows that scanning accuracy, resolution, and grayscale tones meet or exceed the reported manufacturer standards and are consistent across scans. 

All but three images correctly displayed the calibration target as designed. The scanning accuracy compares well to the manufacturer's specifications (Fig \ref{fig:measure_scale}). Tones are also clearly distinguished and consistent across images (Fig \ref{fig:tonal_step_wedge}). The smallest consistently distinguishable feature is on average approximately 46$\mu$m in width, approximately twice the pixel size (Fig \ref{fig:resolution_target}).

In some cases, the scanned images contain features that reflect not ground conditions in the area depicted but artefacts created during exposure, printing, or storage that cannot be remediated (see Figs. \ref{fig:static} to \ref{fig:emulsion}). Awareness of these potential artefacts is particularly important for users who wish to take advantage of computer vision techniques to automate feature identification in the scanned images, because of the potential for confusion with ground features. In many other cases, preservation successfully remediates damage incurred during inappropriate storage in the past so that images are restored to close to their original condition (Fig. \ref{fig:mould}). 

\subsection*{Cost-effectiveness} Automation requires a considerable up-front investment in infrastructure. We estimate that automation becomes cost-effective when scanning more than 2.4 million images, which takes only 65 working weeks if 24/7 production can be sustained. Automation more than halves costs compared to the manual approach when scanning more than 5.0 million images. 

Our benchmark estimates are based on the following assumptions, as well as the productivity estimates we reported above. For the robot-assisted pipeline, we assume for 4 robots and 8 scanners housed in 4 tables, with one additional scanner for manual scanning as required, e.g., for prints that are too fragile for robotic handling. The worker in this scenario dedicates $\sfrac{2}{7}$ of a full-time equivalent position to loading and troubleshooting the robot pipeline, and receives 31,177 GBP in annual compensation \cite{scot2025financial}. For a manual pipeline, we assume one worker operates two scanners. The worker in this scenario works full-time and receives 25,325 GBP in annual compensation \cite{scot2025financial}. We assume the following unit costs of capital equipment: engineered table to house robotic arm and scanner units 80,000 GBP; robotic arm 36,000 GBP; scanner 5,000 GBP; scanner-lid lifting automation 350 GBP. We abstract from rent and energy costs, which are difficult to calculate and compare: the robotic pipeline has a somewhat larger footprint, but the manual pipeline takes much more time. The fixed cost of establishing the robotic pipeline is thus 512,150 GBP and the cost per scan thereafter is 0.0075 GBP, while the fixed cost of establishing the manual pipeline is 10,000 GBP and the cost per scan is 0.22 GBP. With these estimated costs, we can trace out how cost per scan evolves with the total number of scans, yielding the estimates in the previous paragraph (Fig. \ref{fig:break_even}). An important caveat to these estimates is that they are based on UK-specific capital and labour costs.

\subsection*{Discussion} 

The pipeline we describe here exemplifies an approach to collaboration between humans and robots that draws on the strengths of both: the unique abilities of humans to handle complex, unstructured, and varied tasks, and the capacity of robots for unerring, untiring repetition \cite{goldberg2019robots}. The consequence is a massive expansion in feasible output per worker.

The reallocation of tasks creates new, varied, and complex jobs that require high levels of skill and experience. Automating one task in the digitization process---scanning---massively decreases the total cost of digitization, increasing demand for digitization programs and in turn for workers to carry out the tasks that cannot be automated (Figure \ref{fig:overview}). A direct consequence of the development of the automated pipeline is the creation of seven full-time, permanent positions. These workers are engaged across the full range of tasks involved in digitizing the other archives in NCAP's aerial photography collection. 

One might be concerned nonetheless that automating the scanning process displaces greater numbers of less-skilled workers. However, jobs for these workers were, in practice, largely hypothetical. The costs of manual digitization are high enough that NCAP, like most archives, was simply not undertaking digitization at scale. While global data on digitization progress is not available, two of the largest archives worldwide---NCAP, and the United States National Archives and Records Administration (NARA)---had respectively digitized only about 2\% and 0.5\% of their aerial photography holdings before this project began. 

No previous technology, to our knowledge, achieves the same degree of automation for the scanning of photographic prints. Photographic films with fiducial marks and consistent frame spacing---in widespread use since the 1970s---can already be automatically scanned at rates that are comparable to those we achieve here \cite{leicaDSW700V,Dam2004_HighPerformanceScanning}. New techniques are currently being developed to apply edge detection to automate the scanning of older films with uneven frame spacing. 

Lessons from our approach are applicable to many other aerial photography archives. Even subtle differences in archival material may, however, require modifications. In ongoing work, NCAP are applying techniques developed for this project to the Allied Central Interpretation Unit archive dating from the Second World War. Given wartime supply constraints, these images were printed on cheaper, shinier, and more slippery photographic paper, creating new challenges, such as prints that slip rather than lying still on the scanner bed after placement by the robotic arm, resulting in incorrectly oriented scans. Many of the auxiliary techniques we develop here to digitize prints can also be adapted for archives of photography film. 

Beyond the considerable up-front investment in infrastructure, automation also requires preservation and post-processing capacity to match the scanning capacity in order to sustain high production rates. This project was made possible due to research funding that allowed us to undertake scanning at a sufficiently large scale to justify the up-front investment required. Automation will not be individually cost-effective for small archives. New partnerships and collaborations may be needed to completely digitize the entire global patrimony embodied in aerial photography archives. 

Beyond automating scanning, there remains much to do to increase access to aerial photography archives. To use digital images that are not yet georeferenced, users must follow a complex trail of historical documentation to locate each image, for example, on the hand-drawn \emph{sortie plots} that sketched out the coverage of each aerial photography mission, or painstakingly identify the individual features in each image if such documentation has not been preserved. New techniques are also being developed to automate georeferencing and reduce dependency on historical documentation \cite{noda2024machine}, creating footprints for each image that greatly increase the efficiency of searching within an archive for coverage of a given location. Creating digital copies of physical images is, however, a critical first step in increasing access to the many millions of aerial photography images currently stored in archives.

\bibliography{references}

\paragraph*{Data Availability} Data and code to produce statistics and graphs from the manuscript and supplementary materials are available at: https://doi.org/10.7910/DVN/YFSOO6. 

\paragraph*{Acknowledgments}  We gratefully acknowledge funding from Riksbankens Jubileumsfond Infrastructure for Research (IN17-0496:1) and Programme (M20-0044) grants, as well as funding from Jan Wallanders och Tom Hedelius stiftelse samt Tore Browaldhs stiftelse. Pedro Ferreira Rosado, Kasper Madestam, and Diego Valdivia Huaman provided exemplary research assistance. 

\paragraph*{Author Contributions (CRediT)} \emph{Masson}: Conceptualization, Methodology, Investigation, Data Curation, Writing – Original Draft, Supervision, Project Administration, Visualization. 
\emph{Potts}: Conceptualization, Methodology, Supervision, Project Administration. 
\emph{Williams}: Conceptualization, Methodology, Writing – Review \& Editing, Supervision, Project Administration. 
\emph{Berggreen}: Data Curation. 
\emph{McLaren}: Methodology, Data Curation. 
\emph{Martin}: Software, Methodology, Data Curation, Investigation, Validation. 
\emph{Noda}: Data Curation. 
\emph{Nordfors}: Methodology, Software, Validation, Formal Analysis, Investigation, Data Curation, Writing – Review \& Editing, Visualization, Supervision. 
\emph{Ruecroft}: Investigation, Data Curation, Visualization, Writing – Review \& Editing, Validation. 
\emph{Druckenmiller}: Conceptualization, Funding Acquisition, Supervision. 
\emph{Hsiang}: Conceptualization, Funding Acquisition, Supervision. 
\emph{Madestam}: Conceptualization, Funding Acquisition, Supervision, Project Administration, Writing – Review \& Editing. 
\emph{Tompsett}: Conceptualization, Methodology, Funding Acquisition, Supervision, Project Administration, Formal Analysis, Visualization, Writing – Original Draft.

\paragraph{Competing interests} Masson, Potts, Williams, McLaren, Martin, and Ruecroft declare that they are or were employed by Historic Environment Scotland, which curates the National Collection of Aerial Photography and the imagery to which we apply the method described in this manuscript. All other authors declare no financial or non-financial competing interests. 

\clearpage

\clearpage

\appendix

\section*{Supplementary materials}

\setcounter{table}{0}
\setcounter{figure}{0}
\renewcommand{\thetable}{S\arabic{table}}
\renewcommand{\thefigure}{S\arabic{figure}}

%Supplementary figures go here

%Example raw scan
\begin{figure}[ph!]
\caption{Scanned photograph \label{fig:scan}}

\begin{center}
    \includegraphics[width=0.75\textwidth]{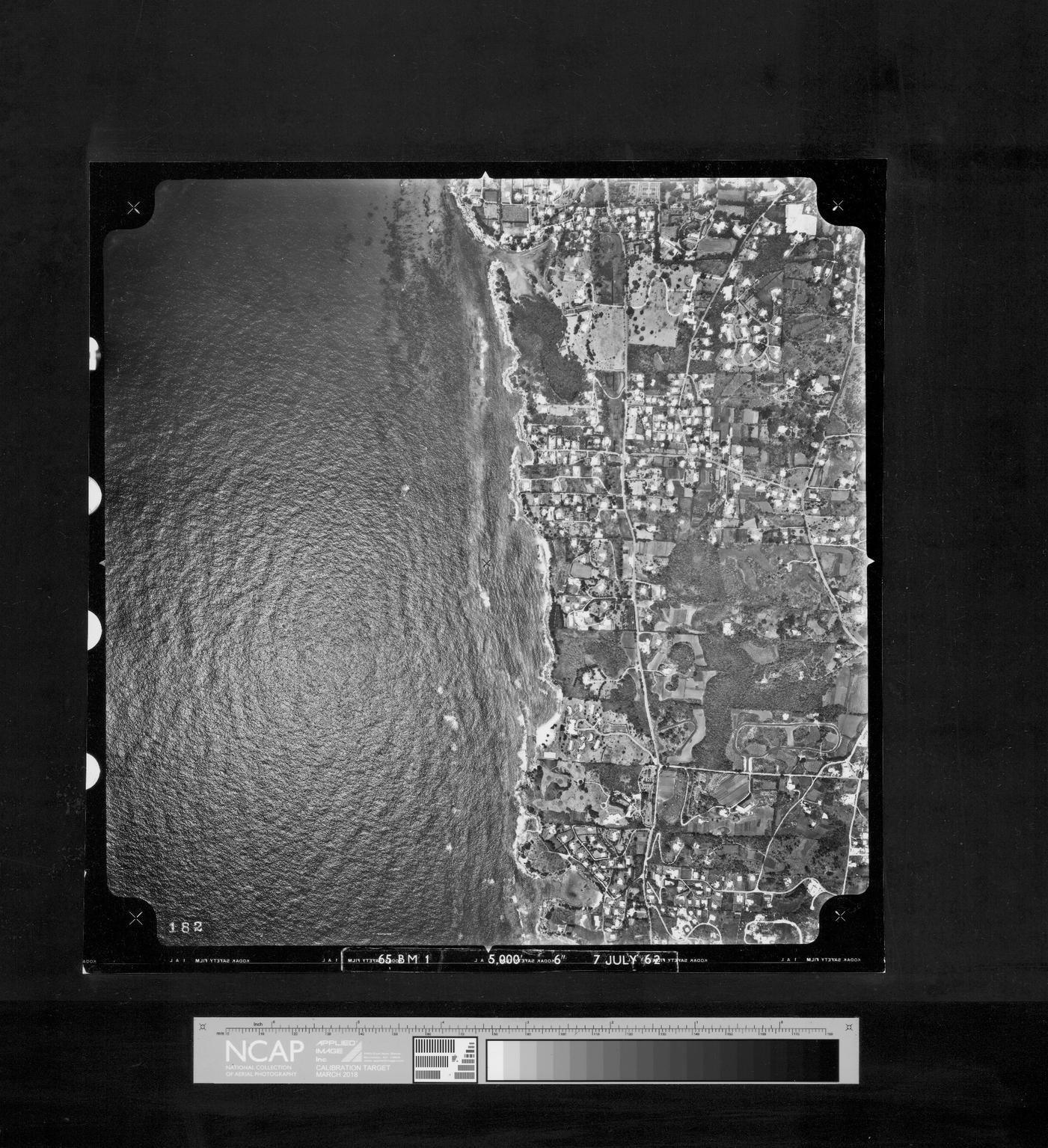}
\end{center}
{\footnotesize \raggedright \emph{Notes} Example of scanned aerial photograph with metadata, fiduciary marks, and calibration target. Not shown at original resolution. \par}
\end{figure}

%Example metadata
\begin{figure}[ph!]
\caption{Metadata\label{fig:metadata}}

\begin{center}
    \includegraphics[width=0.9\textwidth]{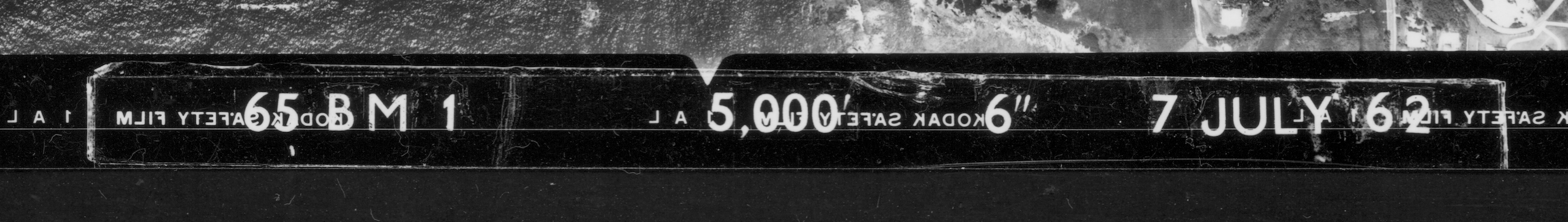}
\end{center}
{\footnotesize \raggedright \emph{Notes} Example of metadata, showing contract reference (65 BM 1), altitude (5000$^\prime$), focal length (6$^{\prime\prime}$, and date (7th July 1962). \par}
\end{figure}

%Altitude
\begin{figure}
\begin{center}

\caption{Altitudes\label{fig:altitudes}}

\begin{subfigure}{0.55\textwidth}
\caption{Weighted distribution\label{fig:altitudes_w}}

\begin{center}
    \includegraphics[width=\textwidth]{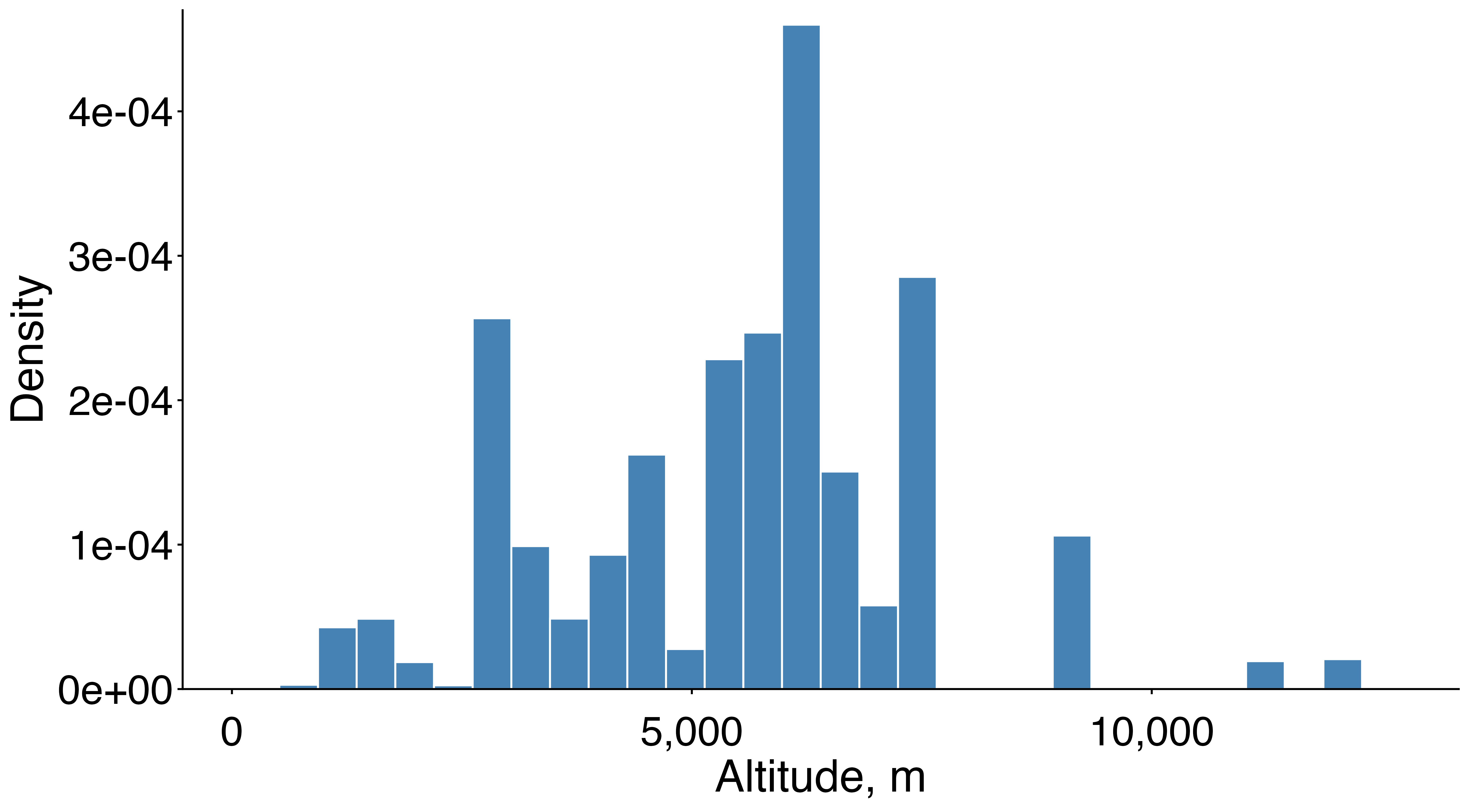}
\end{center}
\end{subfigure}\\
\begin{subfigure}{0.55\textwidth}
\caption{Unweighted distribution\label{fig:altitudes_uw}}

\begin{center}
    \includegraphics[width=\textwidth]{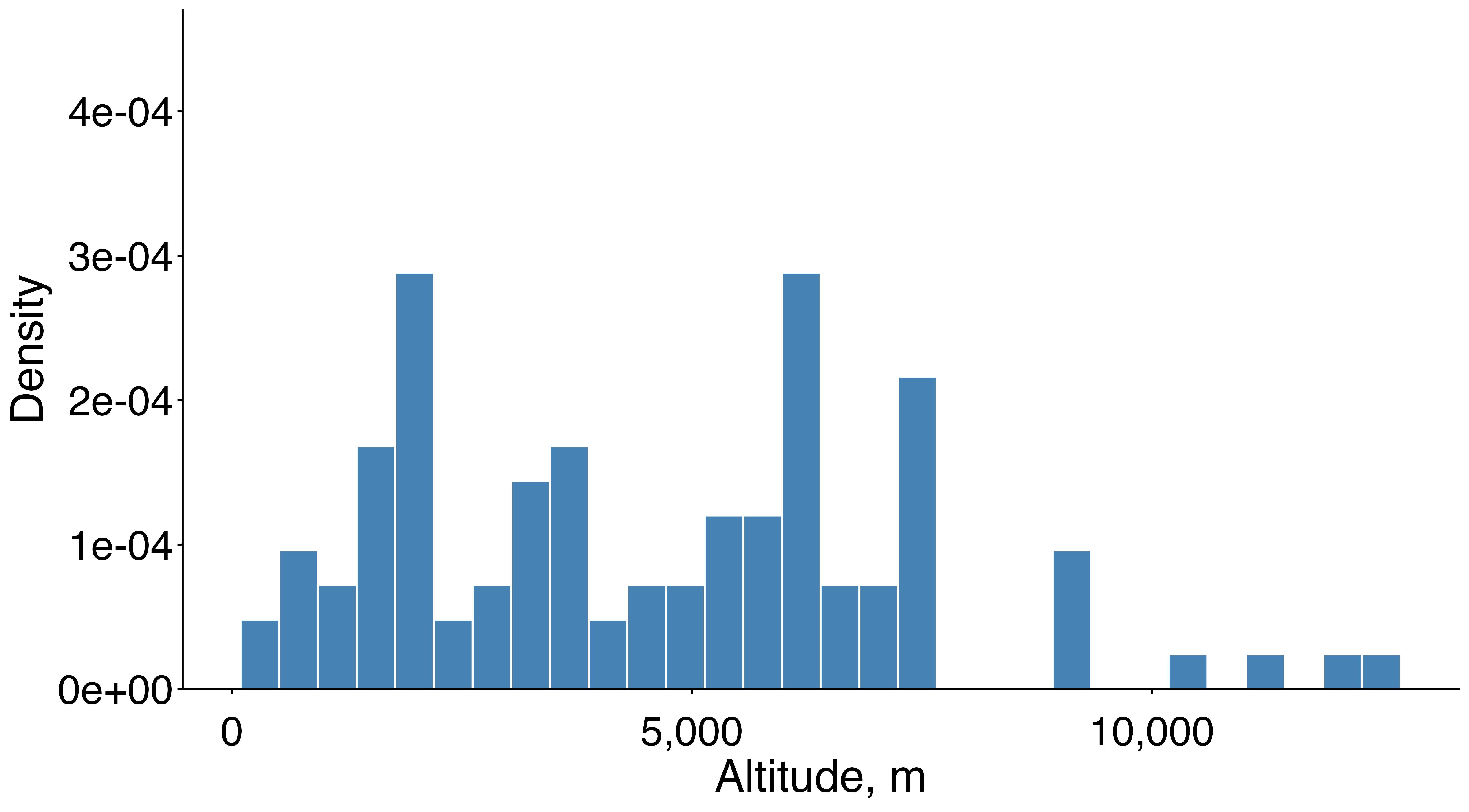}
\end{center}

\end{subfigure}
\end{center}

{\footnotesize \raggedright \emph{Notes}  Distribution of reported altitudes, in m, from validation sample of scanned images. Images are weighted in Fig \ref{fig:altitudes_w} by the share of images in the archive corresponding to each country and decade, or each country when dates were not available at the time of sampling, based on catalogue data. Fig \ref{fig:altitudes_uw} shows the unweighted distribution which is more dispersed because some small surveys have low altitude surveys with only a small number of images. 

\par}

\end{figure}

%Scale
\begin{figure}
\begin{center}

\caption{Scale and ground resolved distance\label{fig:scale}}

\begin{subfigure}{\textwidth}
\caption{Weighted distribution\label{fig:scale_w}}

\begin{center}
    \includegraphics[width=\textwidth]{figures/hist_weighted.png}
\end{center}
\end{subfigure}\\
\begin{subfigure}{\textwidth}
\caption{Unweighted distribution\label{fig:scale_uw}}

\begin{center}
    \includegraphics[width=\textwidth]{figures/hist_dens.png}
\end{center}

\end{subfigure}
\end{center}

{\footnotesize \raggedright \emph{Notes}  Distribution of scales, either reported directly in metadata or approximated from flight altitude and camera focal length. Scales are converted to ground resolved distance---the size of the smallest resolvable object at ground level---assuming i) 10 line pairs per mm \cite{clark1944photographic} and ii) 27 line pairs per mm \cite{light1993national}. Images are weighted in Fig \ref{fig:scale_w} by the share of images in the archive corresponding to each country and decade, or each country when dates were not available at the time of sampling, based on catalogue data. Fig \ref{fig:scale_uw} shows the unweighted distribution which is more dispersed because some small surveys have low altitude surveys, high resolution surveys with only a small number of images. 

\par}

\end{figure}

%Calibration target
\begin{figure}[ph!]
\caption{Calibration target\label{fig:calibration}}

\noindent
\begin{tikzpicture}[>=stealth]
    % Image 
    \node[anchor=north west] (Img) at (0,0)
    {\includegraphics[width=14cm]{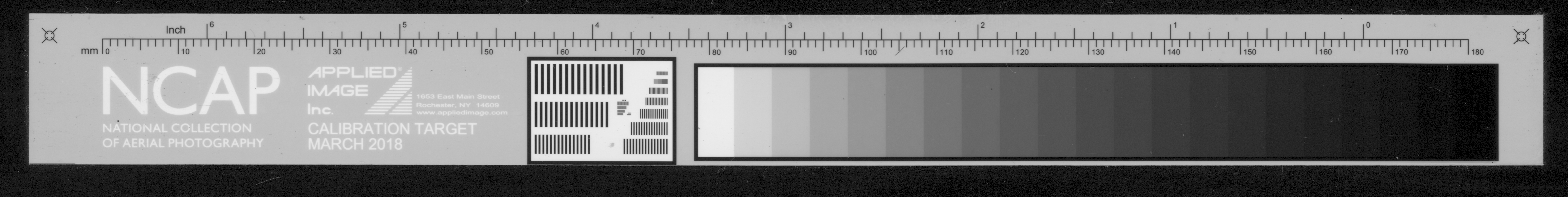}}; 

    \node[draw=black, fill=white, circle, text=black, inner sep=1pt, anchor=north west] at ([xshift=0.6cm, yshift=-0.4cm] Img.north west) {1};

    \node[draw=black, fill=white, circle, text=black, inner sep=1pt, anchor=north west] at ([xshift=4.4cm, yshift=-0.7cm] Img.north west) {2};

    \node[draw=black, fill=white, circle, text=black, inner sep=1pt, anchor=north west] at ([xshift=13.55cm, yshift=-1.2cm] Img.north west) {3};

\end{tikzpicture}

{\footnotesize \raggedright \emph{Notes} Calibration target. Elements comprise a measure scale (1); a resolution target (2); and a tonal step wedge with 21 segments (3). Not shown to scale. \par}
\end{figure}

%Measure scale
\begin{figure}[ph!]
\caption{Validation: measure scale\label{fig:measure_scale}}

\begin{center}
    \includegraphics[height=7cm]{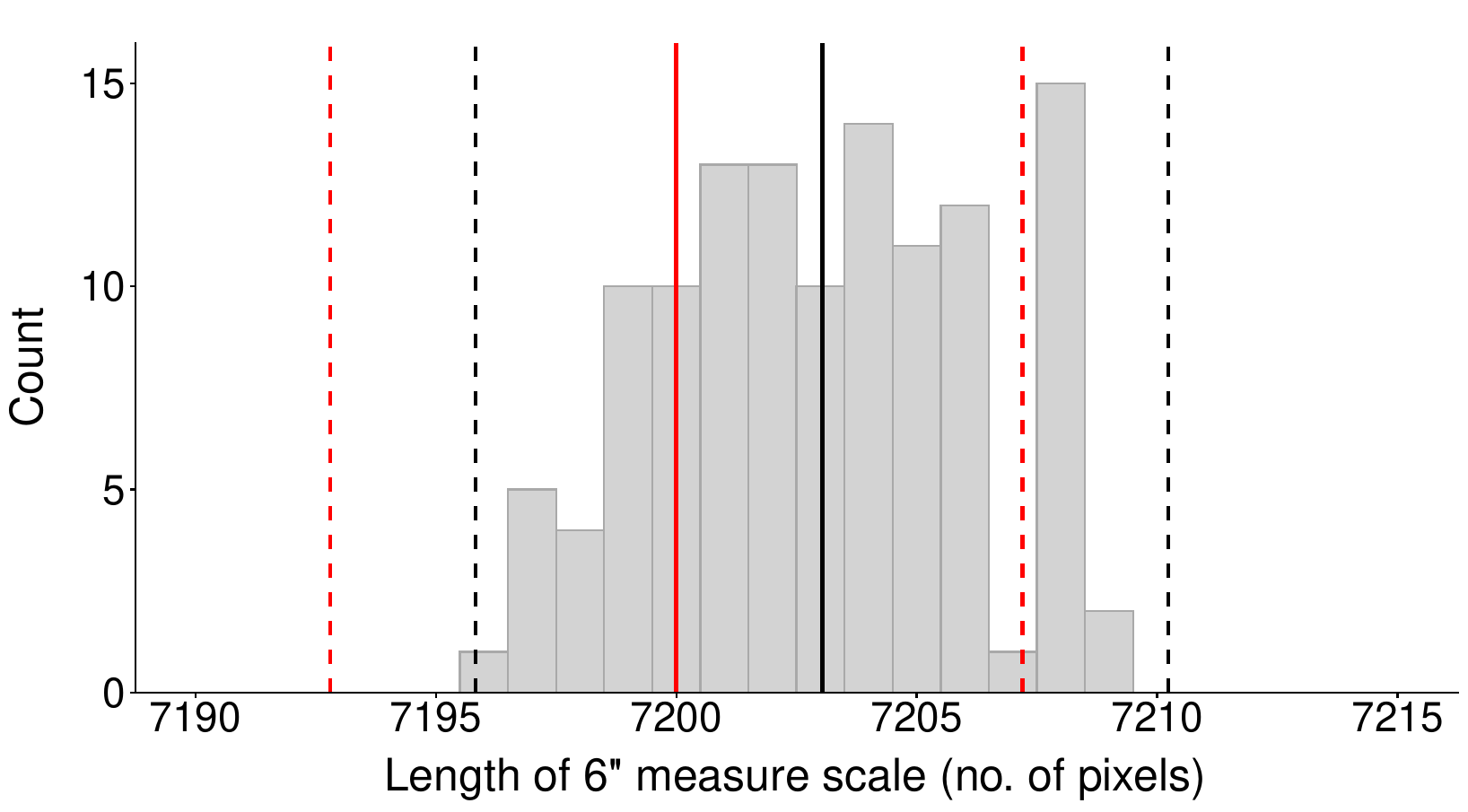}
\end{center}
{\footnotesize \raggedright \emph{Notes}  Distribution of length in pixels of 6'' measure scale in scanned images. The solid red line shows the theoretical length at 1200 dpi, 7200 dpi. The dashed red lines show an interval of $\pm$0.1\% around this value. The vertical black line shows the mean in our validation dataset. That the mean slightly exceeds 7200 dpi shows that our scans are fractionally higher resolution than 1200 dpi. The dashed black lines shown an interval of $\pm$0.1\% around the mean. Noting that labeling of the measure scale is only accurate within 1 to 2 pixels, the results suggest that performance compares well to the manufacturer's reported scanning accuracy of 0.1\% $\pm$ 1 pixel. The clustering of values likely reflects different scanners having slightly different performances. \par}
\end{figure}

%Tonal wedge
\begin{landscape}
\begin{figure}[ph!]
\caption{Validation: tonal step wedge \label{fig:tonal_step_wedge}}

\begin{center}
    \includegraphics[width=21cm]{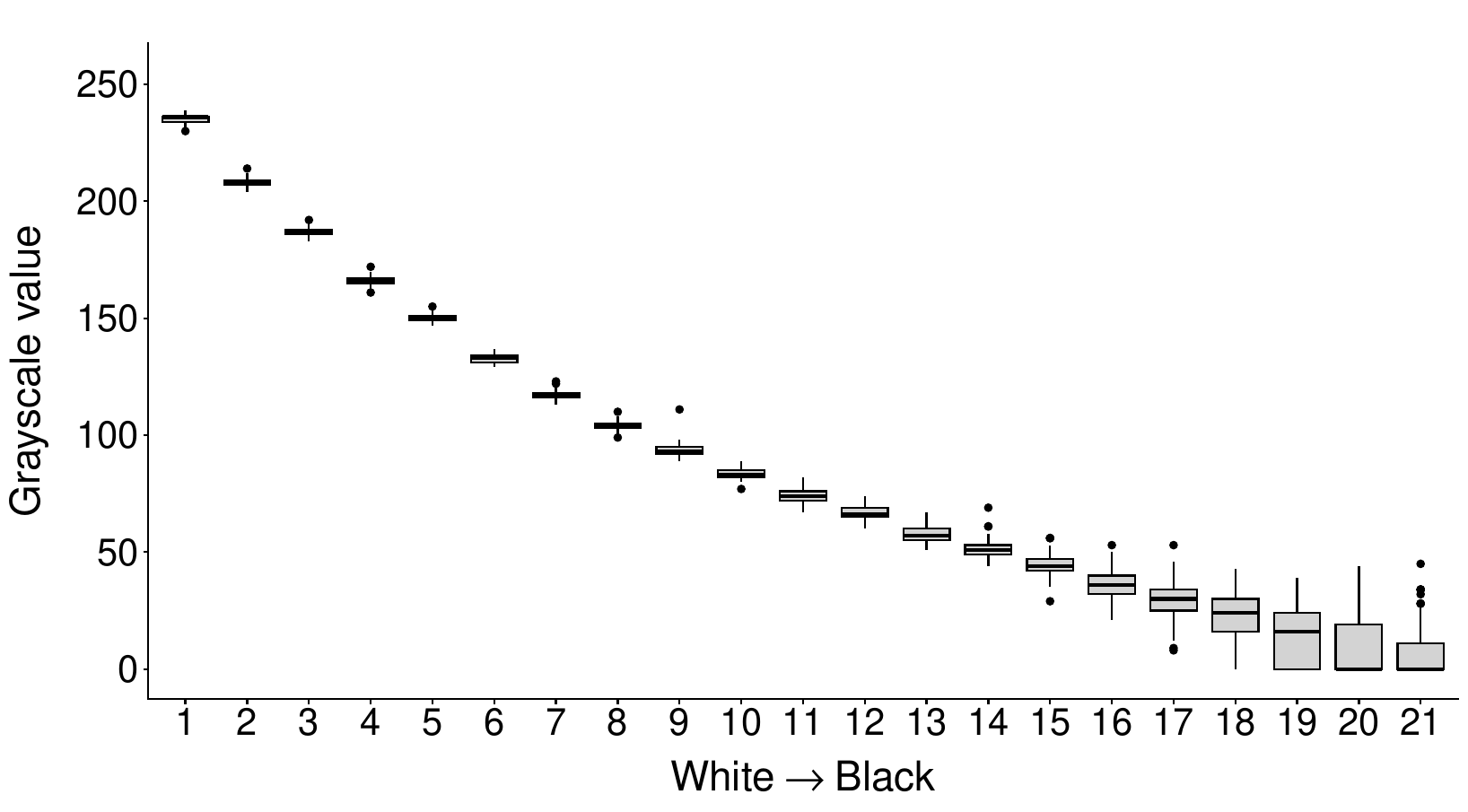}
\end{center}
{\footnotesize \raggedright \emph{Notes} Each boxplot summarizes the range of tonal values for the indicated segment of the tonal step wedge across the images in our validation sample. Boxes shows interquartile range, with the center line in each box showing the median. Whiskers show range as long as it does not exceed 1.5 $\times$ the interquartile range from the median. Outliers (more than 1.5 $\times$ the interquartile range from the median) shown as dots. Except for the very darkest tones, tonal values are well distinguished, with narrow distributions within each segment. 
\par}
\end{figure}
\end{landscape}

%Resolution target
\begin{figure}[ph!]
\caption{Validation: resolution target\label{fig:resolution_target}}

\begin{center}
    \includegraphics[height=7cm]{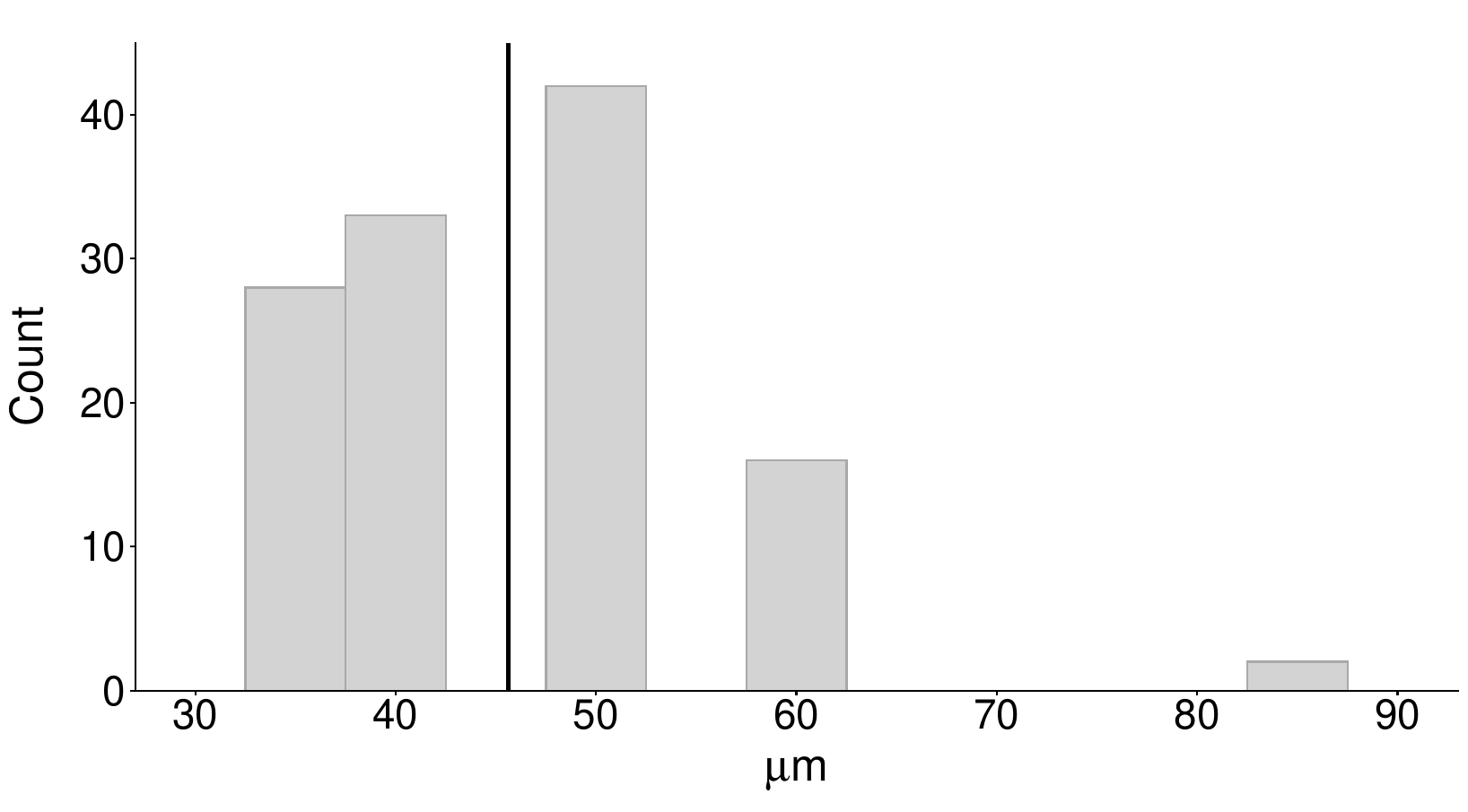}
\end{center}
{\footnotesize \raggedright \emph{Notes} Graph shows the implied distribution of the smallest distinguishable feature, obtained by inspecting the groups of bars of different line widths in the resolution target and identifying the last group of bars that can be clearly distinguished and correctly counted. Black line shows the mean size of the smallest consistently distinguishable line width (45.6~$\mu$m). This performance is consistent with the target resolution of 1200~dpi, where 1 pixel corresponds to approximately 21 $\mu$m. Lines smaller than twice the pixel resolution may not be consistently renderered depending on the position of the pixel raster relative to the lines.
\par}
\end{figure}

%Residual artefacts 

\begin{subfigures}

%Static marks

\begin{landscape}
\begin{figure}[ph!]
\caption{Static marks \label{fig:static}}

\noindent
\begin{tikzpicture}[>=stealth]
    % First (larger) image on the left
    \node[anchor=north west] (leftImg) at (0,0)
    {\includegraphics[height=12cm]{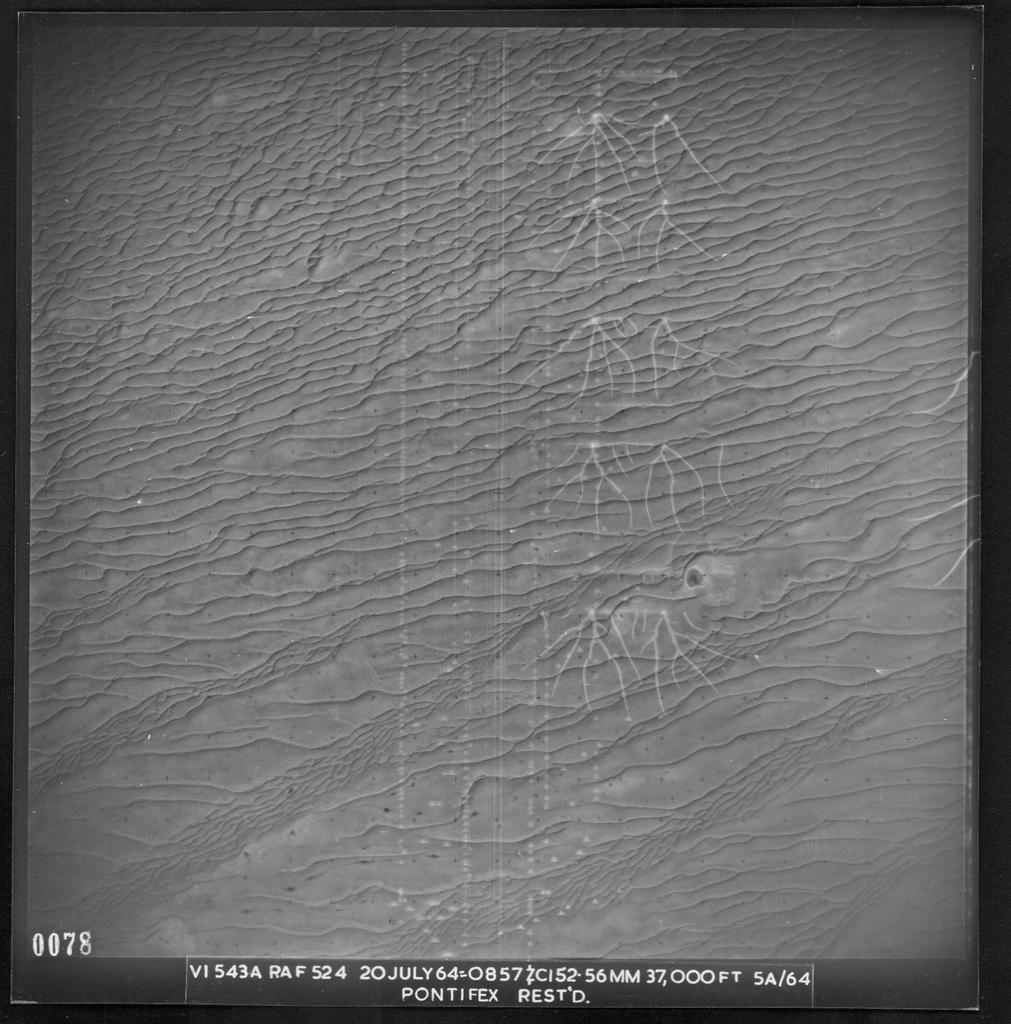}}; 

    \node[draw=black, fill=white, circle, text=black, inner sep=1pt, anchor=north east] at ([xshift=-5.5cm, yshift=-3.5cm] leftImg.north east) {1};

    \node[draw=black, fill=white, circle, text=black, inner sep=1pt, anchor=north east] at ([xshift=-7.5cm, yshift=-8.5cm] leftImg.north east) {2};

    % Second (smaller) image on the right; 
    % note how we position it relative to leftImg.north east
    \node[anchor=north west] (rightImg1) at ([xshift=0.5cm]leftImg.north east)
    {\includegraphics[height=4.5cm]{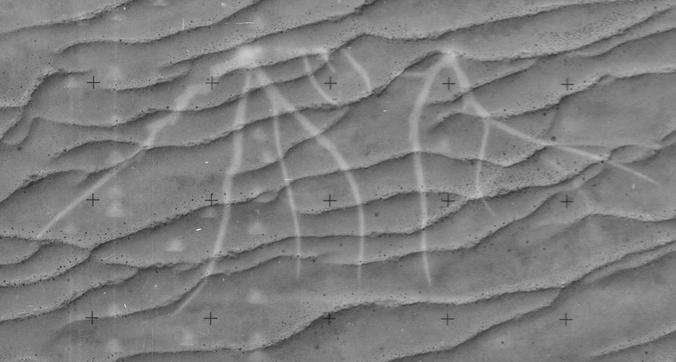}};

    % Label the cropped image
    \node[draw=black, fill=white, circle, text=black, inner sep=1pt, anchor=north west]
        at ([xshift=0.5cm, yshift=-0.5cm] rightImg1.north west) {1};

    % Third (smaller) image on the right; 
    % note how we position it relative to leftImg.north east
    \node[anchor=north west] (rightImg2) at ([yshift=-0.5cm]rightImg1.south west)
    {\includegraphics[height=6cm]{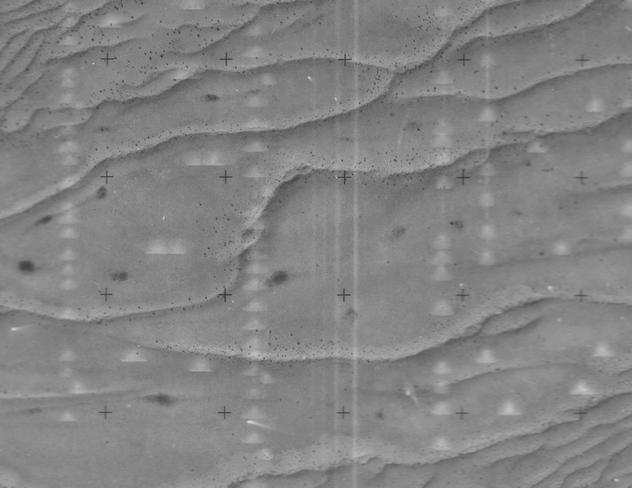}};

    % Label the cropped image
    \node[draw=black, fill=white, circle, text=black, inner sep=1pt, anchor=north west]
        at ([xshift=0.5cm, yshift=-0.5cm] rightImg2.north west) {2};

\end{tikzpicture}

{\footnotesize \raggedright \emph{Notes} Botswana, 1964. Visual artefacts created by static marks on original negatives, typically caused by static electricity generated by winding film in dry conditions.\par}

\end{figure}
\end{landscape}

%Printing error
\begin{landscape}
\begin{figure}[ph!]
\caption{Print error\label{fig:print_error}}

\noindent
\begin{tikzpicture}[>=stealth]
    % First (larger) image on the left
    \node[anchor=north west] (leftImg) at (0,0)
    {\includegraphics[height=12cm]{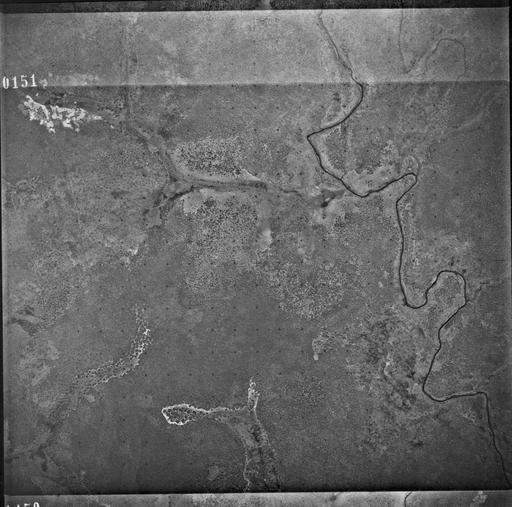}}; 

    \node[draw=black, fill=white, circle, text=black, inner sep=1pt, anchor=north east] at ([xshift=-4.5cm, yshift=-1cm] leftImg.north east) {1};

    \node[draw=black, fill=white, circle, text=black, inner sep=1pt, anchor=north east] at ([xshift=-3cm, yshift=-11cm] leftImg.north east) {2};

    % Second (smaller) image on the right; 
    % note how we position it relative to leftImg.north east
    \node[anchor=north west] (rightImg1) at ([xshift=0.5cm]leftImg.north east)
    {\includegraphics[height=5.5cm]{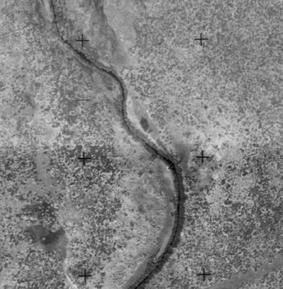}};

    % Label the cropped image
    \node[draw=black, fill=white, circle, text=black, inner sep=1pt, anchor=north west]
        at ([xshift=0.5cm, yshift=-0.5cm] rightImg1.north west) {1};

    % Third (smaller) image on the right; 
    % note how we position it relative to leftImg.north east
    \node[anchor=north west] (rightImg2) at ([yshift=-0.5cm]rightImg1.south west)
    {\includegraphics[height=6cm]{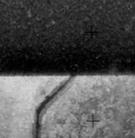}};

    % Label the cropped image
    \node[draw=black, fill=white, circle, text=black, inner sep=1pt, anchor=north west]
        at ([xshift=0.5cm, yshift=-0.5cm] rightImg2.north west) {2};

\end{tikzpicture}

{\footnotesize \raggedright \emph{Notes} Zambia, year uncertain. Printing or exposure error, creating artefacts where shading increases sharply (1) or parts of another image appear (2).\par}

\end{figure}
\end{landscape}

%Processing error
\begin{landscape}
\begin{figure}[ph!]
\caption{Print processing error\label{fig:processing_error}}

\noindent
\begin{tikzpicture}[>=stealth]
    % First (larger) image on the left
    \node[anchor=north west] (leftImg) at (0,0)
    {\includegraphics[height=12cm]{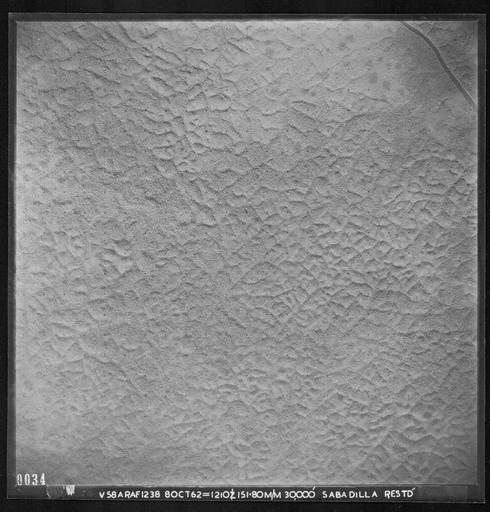}}; 

    \node[draw=black, fill=white, circle, text=black, inner sep=1pt, anchor=north east] at ([xshift=-3.5cm, yshift=-1cm] leftImg.north east) {1};

    % Second (smaller) image on the right; 
    % note how we position it relative to leftImg.north east
    \node[anchor=north west] (rightImg1) at ([xshift=0.5cm]leftImg.north east)
    {\includegraphics[height=7cm]{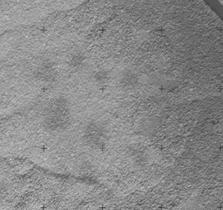}};

    % Label the cropped image
    \node[draw=black, fill=white, circle, text=black, inner sep=1pt, anchor=north west]
        at ([xshift=0.5cm, yshift=-0.5cm] rightImg1.north west) {1};

    % Third (smaller) image on the right; 
    % note how we position it relative to leftImg.north east
    \node[anchor=north west] (rightImg2) at ([yshift=-0.5cm]rightImg1.south west)
    {\includegraphics[height=6cm]{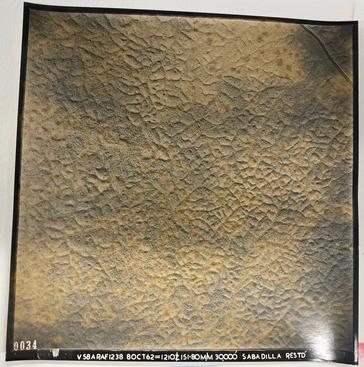}};

\end{tikzpicture}

{\footnotesize \raggedright \emph{Notes} Guyana, 1962. grayscale scanned image shows residual shading inconsistencies as a consequence of processing error during printing. Colour photograph of full print shown bottom right.\par}

\end{figure}
\end{landscape}

%Damage caused by blocking 
\begin{landscape}
\begin{figure}[ph!]
\caption{Blocking damage \label{fig:blocking}}

\noindent
\begin{tikzpicture}[>=stealth]
    % First (larger) image on the left
    \node[anchor=north west] (leftImg) at (0,0)
    {\includegraphics[height=12cm]{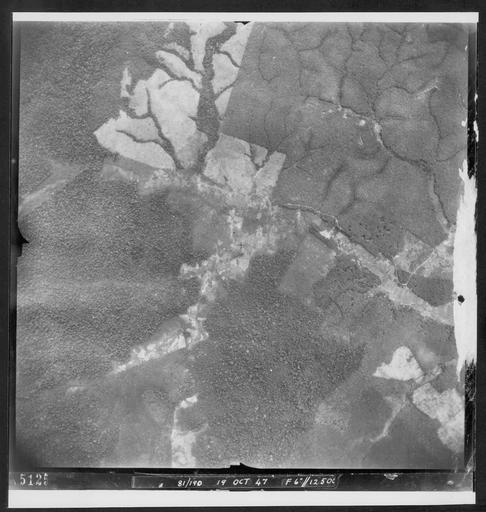}}; 

    \node[draw=black, fill=white, circle, text=black, inner sep=1pt, anchor=north east] at ([xshift=-1.5cm, yshift=-3.5cm] leftImg.north east) {1};

    % Second (smaller) image on the right; 
    % note how we position it relative to leftImg.north east
    \node[anchor=north west] (rightImg) at ([xshift=0.5cm]leftImg.north east)
    {\includegraphics[height=7cm]{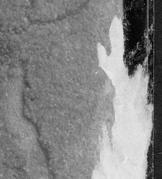}};

    % Label the cropped image
    \node[draw=black, fill=white, circle, text=black, inner sep=1pt, anchor=north west]
        at ([xshift=0.5cm, yshift=-0.5cm] rightImg.north west) {1};

\end{tikzpicture}

{\footnotesize \raggedright \emph{Notes} Malaysia, 1947. Print had been previously blocked with another image and at some point torn apart, leaving paper on the image surface. \par}

\end{figure}
\end{landscape}

%Silver migration
\begin{landscape}
\begin{figure}[ph!]
\caption{Silver migration\label{fig:silver}}

\noindent
\begin{tikzpicture}[>=stealth]
    % First (larger) image on the left
    \node[anchor=north west] (leftImg) at (0,0)
    {\includegraphics[height=12cm]{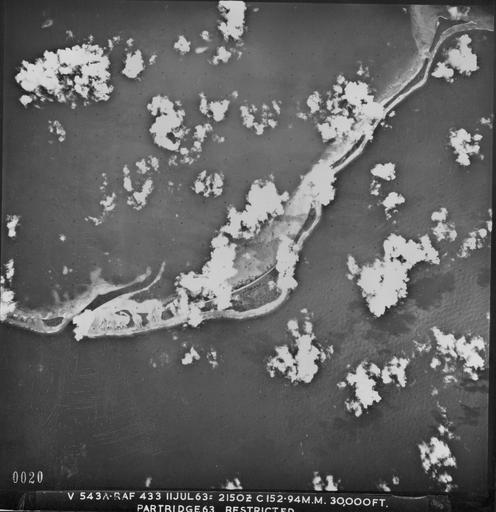}}; 

    \node[draw=black, fill=white, circle, text=black, inner sep=1pt, anchor=north east] at ([xshift=-10.5cm, yshift=-9cm] leftImg.north east) {1};

    % Second (smaller) image on the right; 
    % note how we position it relative to leftImg.north east
    \node[anchor=north west] (rightImg) at ([xshift=0.5cm]leftImg.north east)
    {\includegraphics[height=7cm]{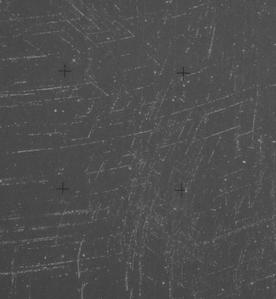}};

    % Label the cropped image
    \node[draw=black, fill=white, circle, text=black, inner sep=1pt, anchor=north west]
        at ([xshift=0.5cm, yshift=-0.5cm] rightImg.north west) {1};

\end{tikzpicture}

{\footnotesize \raggedright \emph{Notes} Solomon Islands, 1963. Scanned image shows residual signs of silver migration in darker areas. \par}

\end{figure}
\end{landscape}

%Annotations
\begin{landscape}
\begin{figure}[ph!]
\caption{Annotations\label{fig:annotations}}

\noindent
\begin{tikzpicture}[>=stealth]
    % First (larger) image on the left
    \node[anchor=north west] (leftImg) at (0,0)
    {\includegraphics[height=12cm]{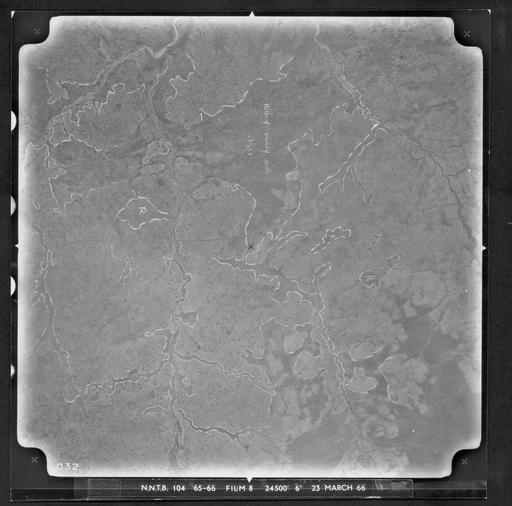}}; 

    \node[draw=black, fill=white, circle, text=black, inner sep=1pt, anchor=north east] at ([xshift=-7cm, yshift=-2cm] leftImg.north east) {1};

    % Second (smaller) image on the right; 
    % note how we position it relative to leftImg.north east
    \node[anchor=north west] (rightImg) at ([xshift=0.5cm]leftImg.north east)
    {\includegraphics[height=7cm]{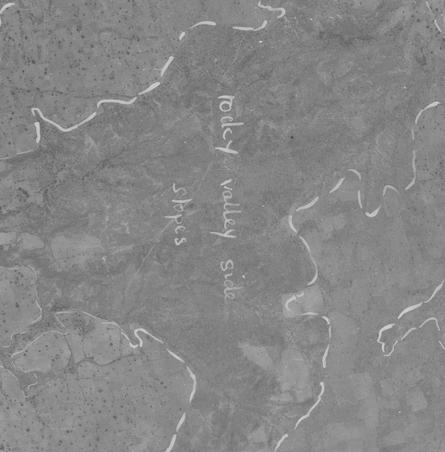}};

    % Label the cropped image
    \node[draw=black, fill=white, circle, text=black, inner sep=1pt, anchor=north west]
        at ([xshift=0.5cm, yshift=-0.5cm] rightImg.north west) {1};

\end{tikzpicture}

{\footnotesize \raggedright \emph{Notes} Nigeria, 1966. Scanned image shows residual historical annotations, noting land cover type and delineating areas with specific land cover type. \par}

\end{figure}
\end{landscape}

%Annotations
\begin{landscape}
\begin{figure}[ph!]
\caption{Damage from adhesives\label{fig:adhesives}}

\noindent
\begin{tikzpicture}[>=stealth]
    % First (larger) image on the left
    \node[anchor=north west] (leftImg) at (0,0)
    {\includegraphics[height=12cm]{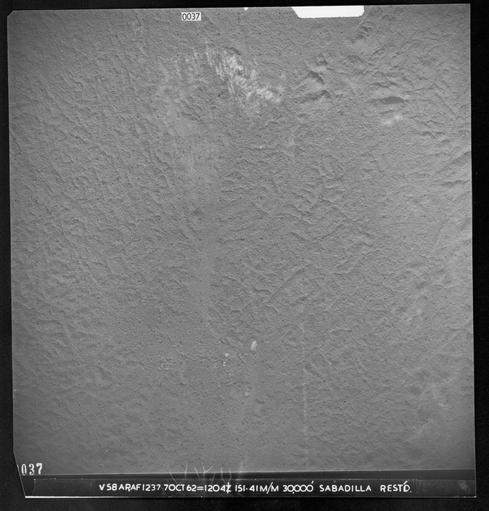}}; 

    \node[draw=black, fill=white, circle, text=black, inner sep=1pt, anchor=north east] at ([xshift=-6.25cm, yshift=-0.5cm] leftImg.north east) {1};

    % Second (smaller) image on the right; 
    % note how we position it relative to leftImg.north east
    \node[anchor=north west] (rightImg) at ([xshift=0.5cm]leftImg.north east)
    {\includegraphics[height=4cm]{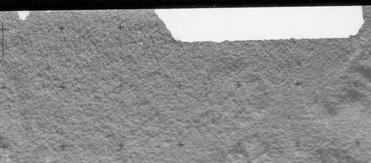}};

    % Label the cropped image
    \node[draw=black, fill=white, circle, text=black, inner sep=1pt, anchor=south west]
        at ([xshift=0.5cm, yshift=0.5cm] rightImg.south west) {1};

\end{tikzpicture}

{\footnotesize \raggedright \emph{Notes} Guyana, 1962.  Tearing of emulsion layer at top of image where adhesive previously applied. \par}

\end{figure}
\end{landscape}

%Emulsion peeling
\begin{landscape}
\begin{figure}[ph!]
\caption{Emulsion peeling\label{fig:emulsion}}

\noindent
\begin{tikzpicture}[>=stealth]
    % First (larger) image on the left
    \node[anchor=north west] (leftImg) at (0,0)
    {\includegraphics[height=12cm]{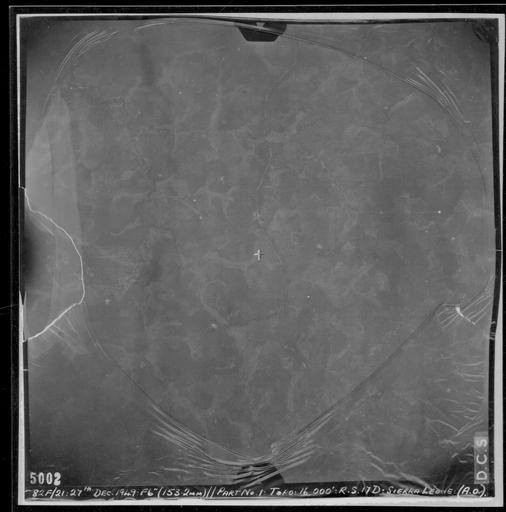}}; 

    \node[draw=black, fill=white, circle, text=black, inner sep=1pt, anchor=north east] at ([xshift=-6cm, yshift=-10cm] leftImg.north east) {1};

    % Second (smaller) image on the right; 
    % note how we position it relative to leftImg.north east
    \node[anchor=north west] (rightImg) at ([xshift=0.5cm]leftImg.north east)
    {\includegraphics[height=7cm]{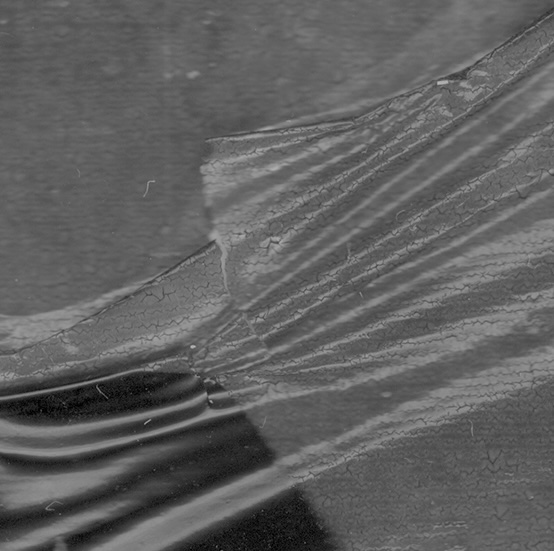}};

    % Label the cropped image
    \node[draw=black, fill=white, circle, text=black, inner sep=1pt, anchor=north west]
        at ([xshift=0.5cm, yshift=-0.5cm] rightImg.north west) {1};

\end{tikzpicture}

{\footnotesize \raggedright \emph{Notes} Sierra Leone, 1949. Emulsion is visibly peeling from paper backing, creating ripple artefacts. Print scanned inside polyester sleeve to prevent additional damage during handling and scanning. \par}

\end{figure}
\end{landscape}

\end{subfigures}

%Mould cleaning 
\begin{landscape}
\begin{figure}[ph!]
\caption{Mould: Before and after cleaning\label{fig:mould}}

\begin{tikzpicture}
  % Top-left image (6cm tall)
  \node[inner sep=0, anchor=north west] (A) at (0,0)
    {\includegraphics[height=6cm]{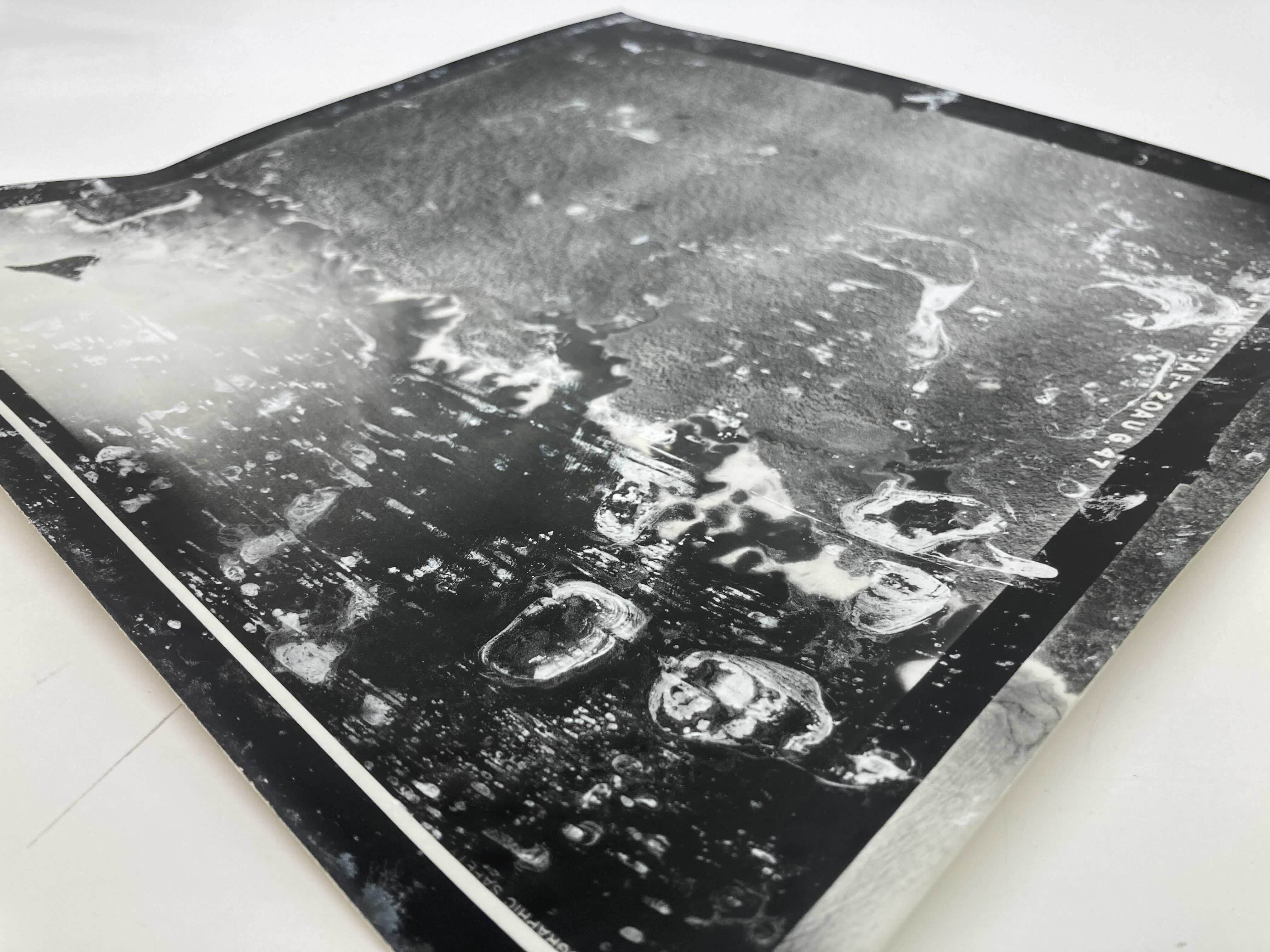}};

  % Second image (6cm tall), 0.5cm below the SW corner of the first
  \node[inner sep=0, anchor=north west] (B) at ([yshift=-0.5cm]A.south west)
    {\includegraphics[height=6cm]{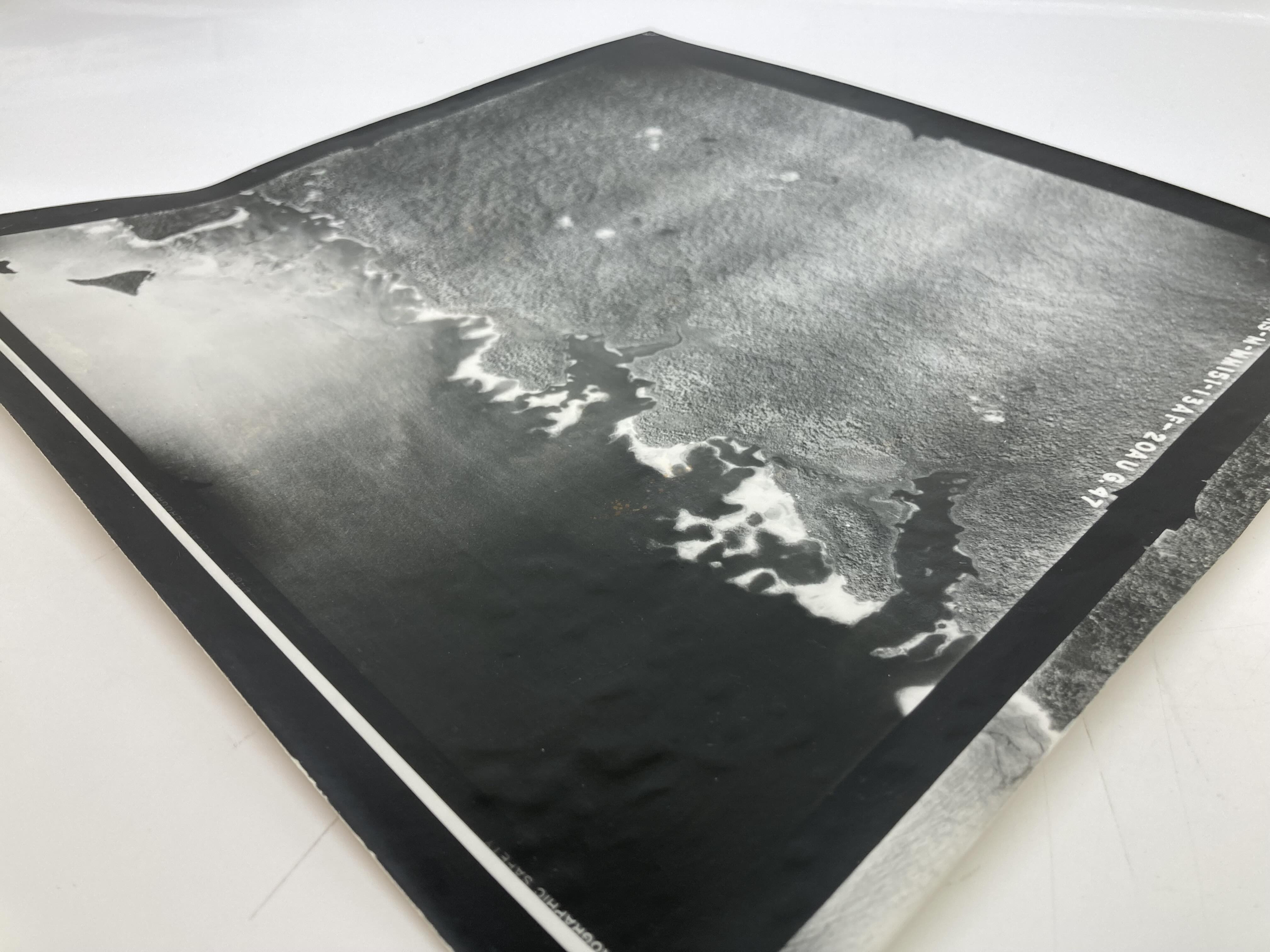}};

  % Right-hand image (12cm tall), NW corner 0.5cm east of the first image
  \node[inner sep=0, anchor=north west] (C) at ([xshift=0.5cm]A.north east)
    {\includegraphics[height=12.5cm]{figures/fig_S1_subfigures/mould_post_NCAP_DOS_338RS_M_0151_0048.jpg}};
\end{tikzpicture}

{\footnotesize \raggedright \emph{Notes} Solomon Islands, 1947. Left-hand images show colour digital photographs of print affected by extensive mould (top) and the same print after cleaning (bottom). Right-hand image shows grayscale scan of the same image, with little residual visible mould. \par}
\end{figure}
\end{landscape}

%Break-even
\begin{figure}[ph!]
\caption{Cost per scan\label{fig:break_even}}

\begin{center}
    \includegraphics[height=7cm]{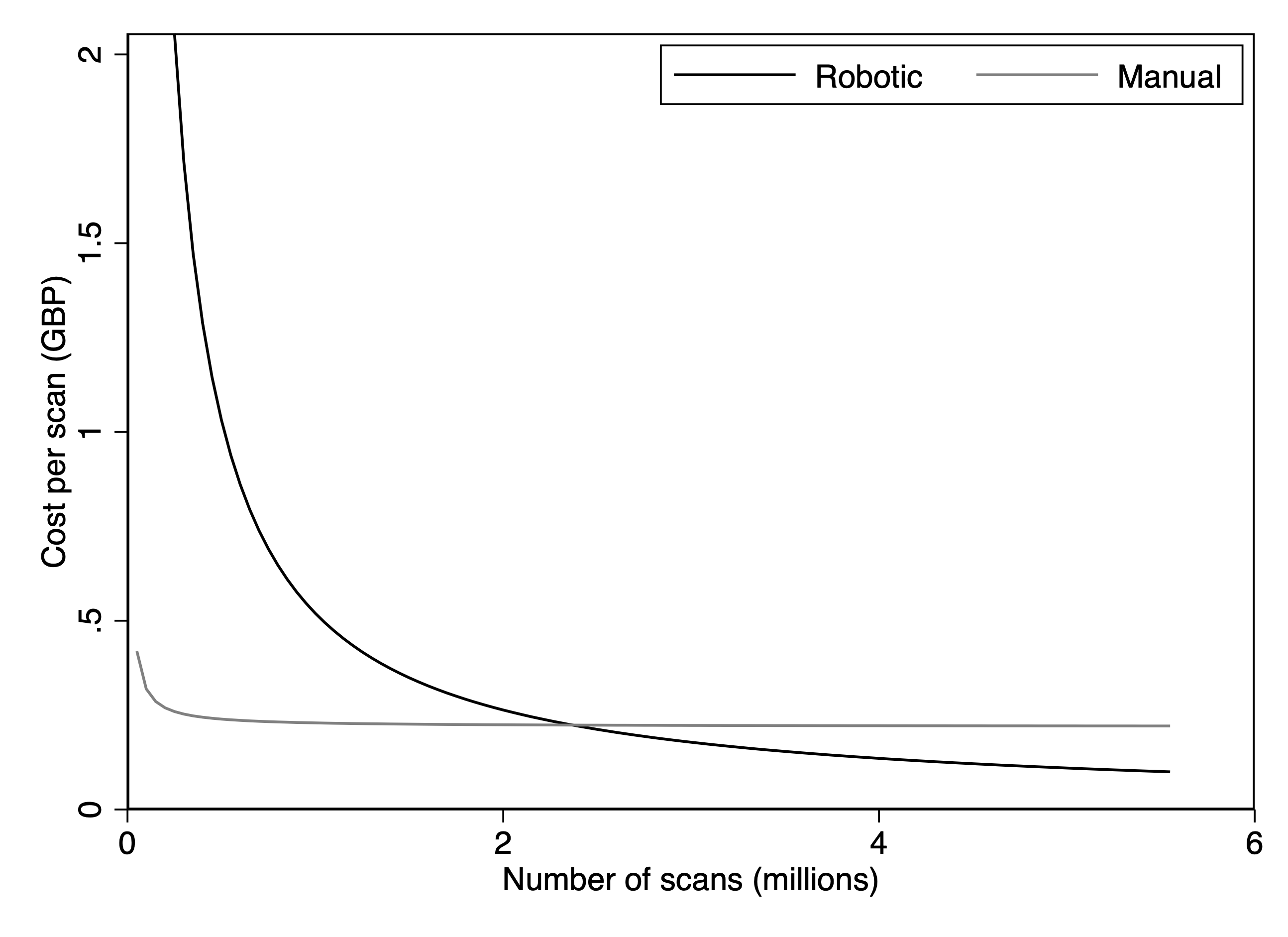}
\end{center}
{\footnotesize \raggedright \emph{Notes} Estimated cost per scan as a function of number of scans produced. Robotic pipeline assumes one human worker, at $\sfrac{2}{7}$ full-time equivalent operating 4 robots and 8 scanners. Manual pipeline assumes one full-time human worker operates 2 scanners. \par}
\end{figure}

\clearpage

%supplementary tables go here

\begin{ThreePartTable}
\begin{center}
\begin{TableNotes} \footnotesize
\item \textit{Notes} No systematic data exists on the extent of aerial photography archives worldwide. To assemble the list in this table, we conducted an internet search and identified 91 archives in 54 countries. The cumulative total holdings of at least 144,756,013 images is a lower bound because not all archives provided online information about the number of images in the archive. We contacted 25 archives by email to request this information and received responses from 6 archives. Dashes indicate unreported data. Archives ordered by continent and country where archive is located. %Image counts include both prints and negatives; not all archives report separate totals. A minority of archives report having digitized a share of their holdings. 
\end{TableNotes}

\begin{longtable}{>{\raggedright\arraybackslash}m{5cm}>{\centering\arraybackslash}p{2cm}>{\centering\arraybackslash}p{2cm}>{\centering\arraybackslash}p{2.5cm}>{\centering\arraybackslash}p{1.2cm}>{\centering\arraybackslash}p{1.2cm}}
\caption{Historical Aerial Photography Archives \label{tab:archives}} \\
\toprule
Name of archive & Continent & Country & No. of images & Start & End  \\
\midrule
\endfirsthead
\multicolumn{6}{c}%
{{\tablename\ \thetable{} -- continued from previous page}} \\
\toprule 
Name of archive & Continent & Country & No. of images & Start & End \\
\midrule
\endhead
\bottomrule \multicolumn{6}{r}{{Continued on next page}} \\
\endfoot
\bottomrule
\insertTableNotes 
\endlastfoot

Egyptian Survey Authority (ESA) & Africa & Egypt & - & - & - \\ \addlinespace
Survey and Mapping Division (SMD) & Africa & Ghana & - & - & - \\ \addlinespace
Regional Centre for Mapping of Resources for Development (RCMRD) & Africa & Kenya & - & - & - \\ \addlinespace
National Space Research and Development Agency (NASRDA) & Africa & Nigeria & - & - & - \\ \addlinespace
Rwanda Land Management and Use Authority & Africa & Rwanda & - & - & - \\ \addlinespace
National Geo-spatial Information (NGI) & Africa & South Africa & - & 1926 & 2008 \\ \addlinespace
The Tanzania Commission for Science and Technology (COSTECH) & Africa & Tanzania & - & - & - \\ \addlinespace
National Agricultural Research Organisation (NARO) & Africa & Uganda & - & - & - \\ \addlinespace
\addlinespace \addlinespace
Lands Department Aerial Photographs & Asia & China & 320,000 & 1924 & 2016 \\ \addlinespace
Survey of India (SOI) & Asia & India & - & - & - \\ \addlinespace
Hebrew University of Jerusalem Aerial Photography Archive & Asia & Israel & 100,000 & 1917 & 1971 \\ \addlinespace
Geospatial Information Authority of Japan (GSI) Collection of of Aerial Photographs Results in the Antarctic & Asia & Japan & - & 1957 & 2003 \\ \addlinespace
The Geospatial Information Authority of Japan (GSI) Aerial Photograph Archive & Asia & Japan & 1,240,000 & 1928 & 2006 \\ \addlinespace
National Geographic Information Institute (NGII) Aerial Photograph Archive & Asia & S.Korea & 894,872 & 1940s & 2009 \\ \addlinespace
Royal Thai Survey Department (RTSD) Aerial Photograph Archive & Asia & Thailand & - & 1952 & 2016 \\ \addlinespace
\addlinespace \addlinespace
Instituto Geográfico Nacional (IGN) Aerial Photograph Archive & C. America & Costa Rica & 123,967 & 1941 & 2000 \\ \addlinespace
Centro Nacional de Registros (CNR) Aerial Photograph Archive & C. America & El Salvador & - & - & - \\ \addlinespace
Instituto Geográfico Nacional de Honduras (IGN) Aerial Photograph Archive & C. America & Honduras & - & - & - \\ \addlinespace
Instituto Nicaragüense de Estudios Territoriales (INETER) Aerial Photograph Archive & C. America & Nicaragua & - & - & - \\ \addlinespace
Instituto Geográfico Nacional Tommy Guardia (IGNTG) Aerial Photograph Archive & C. America & Panama & - & - & - \\ \addlinespace
\addlinespace \addlinespace
Bundesamt für Eich und Vermessungswesen (BEV) Luftbildarchivs & Europe & Austria & 450,000 & 1949 & 2009 \\ \addlinespace
University of Vienna Aerial Archive & Europe & Austria & 120,000 & 1979 & 2006 \\ \addlinespace
Ethiopia Aerial Photographs 1935 & Europe & Belgium & 34,000 & 1930 & 1941 \\ \addlinespace
Department of Lands and Surveys (DLS) aerial photographs & Europe & Cyprus & - & 1945 & - \\ \addlinespace
Land Survey Office of Czech Republic (CUZK) Aerial Survey Photo & Europe & Czech Republic & 770,000 & 1936 & 2010 \\ \addlinespace
Royal Danish Library Aerial Photography Collection & Europe & Denmark & 5,200,000 & 1936 & 1992 \\ \addlinespace
Maaamet's Fotoladu & Europe & Estonia & 260,000 & 1939 & 1993 \\ \addlinespace
National Land Survey of Finland (NLS) Aerial Photographs & Europe & Finland & 1,500,000 & 1931 & 2009 \\ \addlinespace
Institut National de l’Information Géographique (IGN) & Europe & France & 3,400,000 & 1920 & 2005 \\ \addlinespace
Amt für Geoinformation, Vermessungs- und Katasterwesen Luftbilder & Europe & Germany & 107,000 & 1937 & 2003 \\ \addlinespace
Bezirksregierung Köln Geobasis NRW & Europe & Germany & 300,000 & 1954 & 2008 \\ \addlinespace
Landesamt für Digitalisierung, Breitband und Vermessung (LDBV) Luftbilder & Europe & Germany & 1,400,000 & 1941 & 2009 \\ \addlinespace
Landesamt für Geoinformation und Landentwicklung & Europe & Germany & 400,000 & 1934 & 2008 \\ \addlinespace
Landesamt für Vermessung und Geoinformation Thüringen Landesluftbildarchiv & Europe & Germany & 293,000 & 1943 & 2017 \\ \addlinespace
Landesvermessung und Geobasisinformation Brandenburg (LGB) Landesluftbildsammelstelle & Europe & Germany & 150,000 & 1943 & 2007 \\ \addlinespace
The Digital Picture Archives of the Federal Archives & Europe & Germany & 1,300,000 & 1915 & 2005 \\ \addlinespace
Landm{\ae}lingar {\'I}slands Aerial Photo Gallery & Europe & Iceland & 140,000 & 1937 & 2000 \\ \addlinespace
Aerofototeca Nazionale (AFN) & Europe & Italy & 6,000,000 & 1889 & 2009 \\ \addlinespace
IGMI Aerial Photography archive & Europe & Italy & 300,000 & 1927 & 2010 \\ \addlinespace
Latvijas Ģeotelpiskās informācijas aģentūra (LGIA) Orthophoto maps & Europe & Latvia & - & 1994 & 2002 \\ \addlinespace
Kadaster Aerial Photography archive & Europe & Netherlands & 110,000 & 1927 & 2006 \\ \addlinespace
University of Wageningen Library's archive of RAF aerial photographs & Europe & Netherlands & 93,408 & 1943 & 1947 \\ \addlinespace
Kartverkets flyfotoarkiv & Europe & Norway & 1,300,000 & 1935 & 2007 \\ \addlinespace
Head Office of Geodesy and Cartography (GUGiK) database/Military Historical Office (CAW WBH) & Europe & Poland & 1,500,000 & 1951 & 2011 \\ \addlinespace
Centro de Informacao Geoespacial do Exercito (CIGeoE) national archive of aerial photography & Europe & Portugal & 60,000 & 1937 & 1970 \\ \addlinespace
Geodetic Institute of Slovenia (GIS) archives of aerial photos & Europe & Slovenia & - & - & - \\ \addlinespace
Instituto Geográfico Nacional (IGN) Fototeca Digital del CNIG & Europe & Spain & 500,000 & 1945 & 2006 \\ \addlinespace
Ministerio de Agricultura, Pesca y Alimentación Archivo de Fotografías Aéreas & Europe & Spain & 14,560 & 1976 & 1977 \\ \addlinespace
TRACASA Archive of Navarra’s Government & Europe & Spain & 150,000 & 1927 & 2006 \\ \addlinespace
Lantmäteriets historiska ortofoton & Europe & Sweden & 1,000,000 & 1949 & 2005 \\ \addlinespace
Swisstopo’s Aerial images collection & Europe & Switzerland & 500,000 & 1926 & 2010 \\ \addlinespace
Britain from Above Aerofilms collection & Europe & U.K. & 1,260,000 & 1919 & 2006 \\ \addlinespace
Cambridge University Collection of Aerial Photography (CUCAP) & Europe & U.K. & 500,000 & 1945 & 2009 \\ \addlinespace
Central Register of Aerial Photographs for Wales (CRAPW) & Europe & U.K. & 350,000 & 1940 & - \\ \addlinespace
Imperial War Museum's Air Ministry First World War Official Aerial Photographs Collection & Europe & U.K. & 150,000 & 1914 & 1918 \\ \addlinespace
National Collection of Aerial Photography (NCAP) & Europe & U.K. & 30,000,000 & 1924 & 2010 \\ \addlinespace
Royal Commission on the Ancient and Historical Monuments of Wales (RCAHMW) & Europe & U.K. & 2,000,000 & 1920 & 2009 \\ \addlinespace
The Commonwealth and African Aerial Photograph Archive & Europe & U.K. & 1,500,000 & 1940s & 1970s \\ \addlinespace
The Historic England Archive & Europe & U.K. & 6,000,000 & 1919 & 2010 \\ \addlinespace
The Public Records of Northern Ireland (PRONI) archive of aerial photography & Europe & U.K. & 1,500 & 1940 & 2007 \\ \addlinespace
\addlinespace \addlinespace
McMaster University Aerial Photography Collection & N. America & Canada & 17,954 & 1919 & 2000 \\ \addlinespace
National Air Photo Library (NAPL) & N. America & Canada & 5,000,000 & 1920 & 2010 \\ \addlinespace
Western University Air Photo Collection & N. America & Canada & 60,000 & 1922 & 2001 \\ \addlinespace
Wilfrid Laurier University Laurier Military History Archive First Canadian Army Air Photo collection & N. America & Canada & 130,000 & 1942 & 1959 \\ \addlinespace
The National Collection (INEGI) & N. America & Mexico & 978,097 & 1967 & 2010 \\ \addlinespace
Alaska State Archives Alaskan aerial photographs & N. America & USA & 1,924 & 1940 & 1990 \\ \addlinespace
Florida Department of Transportation (FDOT) Aerial Photography Archive & N. America & USA & 700,000 & 1940 & 2003 \\ \addlinespace
Instituto Geográfico Nacional (IGN) Aerial Photograph Archive & N. America & USA & 17,280 & 1960 & 2001 \\ \addlinespace
The Digital Library of Georgia Aerial Photography Collection & N. America & USA & 230,000 & 1938 & 1999 \\ \addlinespace
The National Archives Aerial Photography Archive & N. America & USA & 35,000,000 & 1918 & 2011 \\ \addlinespace
The University of Arizona Aerial photograph collection & N. America & USA & - & 1948 & 2001 \\ \addlinespace
U.S. Geological Survey (USGS) EROS Archive & N. America & USA & 6,500,000 & 1937 & 2014 \\ \addlinespace
UCSB Library collection of aerial photography & N. America & USA & 2,500,000 & 1924 & 2011 \\ \addlinespace
University of Alabama Aerial photography collection & N. America & USA & 75,000 & 1930 & 1999 \\ \addlinespace
University of Florida Aerial Photography: Florida collection & N. America & USA & 160,000 & 1937 & 1990 \\ \addlinespace
Vintage Aerial Photograph Collection & N. America & USA & 18,000,000 & 1960 & 2000 \\ \addlinespace
William P. MacConnell Aerial Photograph Collection & N. America & USA & 24,000 & 1951 & 2000 \\ \addlinespace
York County Archives Aerial Photographs & N. America & USA & - & 1955 & 1982 \\ \addlinespace
\addlinespace \addlinespace
Ministerio de Desarrollo Urbano y Transporte (MDUyT) Fotografias Aereas & S. America & Argentina & - & 1929 & 2002 \\ \addlinespace
Instituto Geográfico Militar (IGM) Aerial Photography Archive & S. America & Bolivia & - & - & - \\ \addlinespace
Instituto Brasileiro de Geografia e Estatística (IBGE) Aerial Photography Archive & S. America & Brasil & 400,000 & 1942 & 2010 \\ \addlinespace
Instituto Geográfico Militar de Chile (IGM) Aerial Photography Archive & S. America & Chile & 95,000 & 1944 & 1961 \\ \addlinespace
Instituto Geográfico Agustín Codazzi (IGAC) Aerial Photograph Collection & S. America & Colombia & 116,037 & 1950 & 2007 \\ \addlinespace
Instituto Geográfico Militar del Ecuador (IGM) Aerial Photography Archive & S. America & Ecuador & - & 1942 & - \\ \addlinespace
Instituto Nacional de Estadística (INE) Aerial Photography Archive & S. America & Paraguay & - & - & - \\ \addlinespace
Instituto Geográfico Nacional (IDEP) Aerial Photography Archive & S. America & Peru & - & - & - \\ \addlinespace
Instituto Geográfico Militar (IGM) Aerial Photography Archive & S. America & Uruguay & 67,944 & 1929 & 1987 \\ \addlinespace
\addlinespace \addlinespace
Historical Aerial Photography collection (HAP) & Oceania & Australia & 1,400,000 & 1928 & 1997 \\ \addlinespace
National Archives of Australia Aerial Photographs (NAA) & Oceania & Australia & - & 1916 & 1997 \\ \addlinespace
National Library of Australia Aerial photographs collection & Oceania & Australia & 790,470 & 1928 & 1990 \\ \addlinespace
LINZ Aerial Imagery Archive & Oceania & New Zealand & 700,000 & 1936 & 2008 \\ \addlinespace
\midrule
Total number of reported images &  &  & 144,756,013 &  &  \\

\end{longtable}
\end{center}
\end{ThreePartTable}

\begin{table}[h]

\caption{Identifying information \label{tab:identifiers}}
\small
\centering

\begin{tabular}{@{}p{0.35\textwidth} p{0.65\textwidth}@{}}
\toprule
\textbf{DOS-contracted imagery} &
Contract number/two-letter ISO country code/sequential mission (film) number, i.e., 4/BC/0056 is DOS Contract 4, Bechuanaland (Botswana after 1960), film 56. \\
\addlinespace
\textbf{British RAF or Royal Navy imagery} &
Unit/service/sequential mission number, e.g., 58 Squadron RAF, mission 456 is shown as \mbox{58/RAF/0456}. \\
\addlinespace
\textbf{Locally contracted commercial survey imagery} &
The standard varies but is usually company/country/year/sequential film number, i.e., HSL/GH/64/0034 is Hunting Surveys Limited photography of Ghana in 1964, film 34. \\
\addlinespace
\textbf{U.S. Army Air Force imagery} &
Labels were created locally and not standardized until the mid-1950s, after which they resemble the British RAF standard. \\
\bottomrule
\end{tabular}

\end{table}

\begin{table}[h]
\caption{Rates of preservation issues \label{tab:preservation}}
\small
\centering
\begin{tabular}{L{8cm} C{2cm} C{2cm}}
\toprule
\textbf{Issue} & \textbf{Number of boxes} & \textbf{\% of total} \\
\midrule
Mould treatment & 250 & 1.5\% \\
Blocked prints separated & 26 & 0.2\% \\
Prints requiring cleaning & 2825 & 17\% \\
Tape removal & 579 & 3.5\% \\
Curled prints humidified & 2823 & 17\% \\
Ripped or torn prints & 259 & 2.0\% \\
Emulsion Peeling & 291 & 1.8\% \\ 
\addlinespace
Boxes with any interventive preservation performed & 6802 & 41\% \\ \midrule
Total DOS project boxes containing prints & 16634 & 100\% \\
\bottomrule
\end{tabular}

\begin{tablenotes} 
\footnotesize
\item \textit{Note:} Total number of boxes with any interventive preservation performed is lower than the sum of boxes with particular preservation issues, because many boxes have multiple issues.
\end{tablenotes}
\end{table}

\end{document}